\documentclass[prd,superscriptaddress,amsfonts,amssymb,amsmath,showpacs,onecolumn]{revtex4-2}
\usepackage{bm}
\usepackage{amsfonts}
\usepackage{graphicx}
\usepackage{amsmath}
\usepackage{palatino}
\usepackage{mathpazo}
\usepackage{textcomp}
\usepackage[utf8]{inputenc}
\usepackage[T1]{fontenc}
\usepackage{float}
\usepackage{booktabs}
\usepackage{dcolumn}
\usepackage{multirow}
\usepackage{hyperref}
\hypersetup{colorlinks,citecolor=blue}
\usepackage{amsmath}
\usepackage{xcolor}
\usepackage{orcidlink}
\usepackage[caption=false]{subfig}
\usepackage{commath}
\usepackage{amssymb}

\usepackage{esdiff}
\usepackage{mathtools}
\usepackage{ifthen}
\usepackage{mathtools, amssymb} % !!! Carefull, incompatibility found here, please see below

\usepackage{siunitx}
\usepackage{tikz}
\usetikzlibrary{hobby} % for ..
\usetikzlibrary{arrows.meta} % to control arrow size
\tikzset{>={Latex[length=4,width=4]}} % for LaTeX arrow head
\usetikzlibrary{calc,intersections,decorations.markings}
\usepackage{siunitx}
\usepackage{xcolor} % for colored text

\colorlet{mylightblue}{blue!20}
\colorlet{myblue}{blue!50!black}
\colorlet{mydarkblue}{blue!30!black}
\colorlet{mylightred}{red!10}
\colorlet{myred}{red!50!black}
\colorlet{mydarkred}{red!60!black}
\colorlet{mydarkgreen}{green!30!black}

\tikzset{
  midarr/.style={decoration={markings,mark=at position #1 with {\arrow{stealth}}},postaction={decorate}},
  midarr/.default=0.5
}

\def\jnl@style{\it}
\def\aaref@jnl#1{{\jnl@style#1}}

\def\aaref@jnl#1{{\jnl@style#1}}

\def\aj{\aaref@jnl{AJ}}                   % Astronomical Journal
\def\apj{\aaref@jnl{ApJ}}                 % Astrophysical Journal
\def\apjl{\aaref@jnl{ApJ}}                % Astrophysical Journal, Letters
\def\apjs{\aaref@jnl{ApJS}}               % Astrophysical Journal, Supplement
\def\apss{\aaref@jnl{Ap\&SS}}             % Astrophysics and Space Science
\def\aap{\aaref@jnl{A\&A}}                % Astronomy and Astrophysics
\def\aapr{\aaref@jnl{A\&A~Rev.}}          % Astronomy and Astrophysics Reviews
\def\aaps{\aaref@jnl{A\&AS}}              % Astronomy and Astrophysics, Supplement
\def\mnras{\aaref@jnl{Mon.~Not.~Roy.~Astron.~Soc.}}             % Monthly Notices of the RAS
\def\prd{\aaref@jnl{Phys.~Rev.~D}}        % Physical Review D
\def\prc{\aaref@jnl{Phys.~Rev.~C}}  % Physical Review C
\def\prl{\aaref@jnl{Phys.~Rev.~Lett.}}    % Physical Review Letters
\def\qjras{\aaref@jnl{QJRAS}}             % Quarterly Journal of the RAS
\def\skytel{\aaref@jnl{S\&T}}             % Sky and Telescope
\def\ssr{\aaref@jnl{Space~Sci.~Rev.}}     % Space Science Reviews
\def\zap{\aaref@jnl{ZAp}}                 % Zeitschrift fuer Astrophysik
\def\nat{\aaref@jnl{Nature}}              % Nature
\def\aplett{\aaref@jnl{Astrophys.~Lett.}} % Astrophysics Letters
\def\apspr{\aaref@jnl{Astrophys.~Space~Phys.~Res.}} % Astrophysics Space Physics Research
\def\physrep{\aaref@jnl{Phys.~Rep.}}      % Physics Reports
\def\physscr{\aaref@jnl{Phys.~Scr}}       % Physica Scripta
\def\commat{\aaref@jnl{Comm.~Math.~Phys.}}              % Communications in Mathematical Physics
\def\science{\aaref@jnl{Science}}               % Science
\def\cqg{\aaref@jnl{Classical Quant.~Grav.}}            % Classical and Quantum Gravity
\def\jpcs{\aaref@jnl{JPCS}}                                     % Journal of Physics Conference Series
\def\ijmpd{\aaref@jnl{Int.~J.~Mod.~Phys.~D}}                    % International Journal of Modern Physics D
\def\grg{\aaref@jnl{Gen.~Relat.~Gravit.}}               % General Relativity and Gravitation
\def\rpp{\aaref@jnl{Rep.~Prog.~Phys.}}          % Reports on Progress in Physics
\def\npa{\aaref@jnl{Nucl.~Phys.~A}}        % Nuclear Physics A
\def\lrr{\aaref@jnl{Living Rev.~Rel.}}                   % Living reviews in relativity
\def\jcap{\aaref@jnl{J.~Cosmology Astropart.~Phys.}}    % Journal of cosmology and astroparticle physics
\def\rmp{\aaref@jnl{Rev.~Mod.~Phys.}}   %Reviews of modern physics
\def\epjc{\aaref@jnl{Eur.~Phys.~J.~C}}

\hypersetup{linkcolor=blue}

\allowdisplaybreaks[1]
\begin{document}

\color{black}       %% For one column

\title{Compact stars admitting Finch-Skea symmetry in the presence of various matter fields}

\author{Oleksii Sokoliuk\orcidlink{0000-0003-4503-7272}}
\email{oleksii.sokoliuk@mao.kiev.ua}
\affiliation{Main Astronomical Observatory of the NAS of Ukraine (MAO NASU),\\
Kyiv, 03143, Ukraine}
\affiliation{Astronomical Observatory, Taras Shevchenko National University of Kyiv, \\
3 Observatorna
St., 04053 Kyiv, Ukraine}

\author{Alexander Baransky\orcidlink{0000-0002-9808-1990}}
\email{abransky@ukr.net}
\affiliation{Astronomical Observatory, Taras Shevchenko National University of Kyiv, \\
3 Observatorna
St., 04053 Kyiv, Ukraine}
\author{P.K. Sahoo\orcidlink{0000-0003-2130-8832}}
\email{pksahoo@hyderabad.bits-pilani.ac.in}
\affiliation{Department of Mathematics, Birla Institute of Technology and
Science-Pilani,\\ Hyderabad Campus, Hyderabad-500078, India.}

%%%%%%%%%%%%%%%%%%%%%%%%%%%%%%%%%%%%  DATE  %%%%%%%%%%%%%%%%%%%%%%%%%%%%%%%%%%%%
\date{\today}
\begin{abstract}

In the present article authors investigate the anisotropic stellar solutions admitting Finch-Skea symmetry (viable and non-singular metric potentials) in the presence of some exotic matter fields, such as: Bose-Einstein Condensate (BEC) dark matter, Kalb-Ramond fully anisotropic rank-2 tensor field from the low-energy string theory effective action, gauge field imposing $U(1)$ symmetry. Interior spacetime is matched with both Schwarzchild and Reissner-N\"ordstrom vacuum spacetimes for BEC, KB and gauge fields respectively. Further, we have studied energy conditions, Equation of State (EoS), radial derivatives of energy density and anisotropic pressures, Tolman-Oppenheimer-Volkoff equilibrium condition, relativistic adiabatic index, sound speed and surface redshift. Most of the aforementioned conditions were satisfied and therefore solutions derived in the current study lie in the physically acceptable regime.

\textbf{Keywords:} Compact stars, Finch-Skea symmetry, Energy conditions, Tolman-Oppenheimer-Volkoff equilibrium condition, sound speed
\end{abstract}

\maketitle
\section{Introduction} \label{sec:1}
Generally, compact stars are the final stage of the stellar evolution, that could be formed because of the radial pressure from core nuclear fusion that is bigger that gravitational forces. Gravitational compact objects include many various objects, such as white dwarfs, neutron stars, black holes and naked singularities. Moreover, compact star could also include more exotic objects, such as strange stars (made of strange quarks), gravitational condensate stars (non-singular three-layer alternative to the black hole) etc. Such compact objects usually were studied in the General Theory of Relativity (further - GR), but there are also present a couple of works in the modified gravity formalism (for example, in the $f(\mathcal{R})$ gravity \cite{Pretel:2022plg,Jimenez:2021wik,Numajiri:2021nsc,Sharif:2021emv,Abbas:2021bid}, in the $f(\mathcal{R},\mathcal{T})$ gravity \cite{Bhar:2021uqr,Ahmed:2021fav,Kumar:2021vqa,Bhar:2021iog,Pretel:2021kgl} and finally in the teleparallel modified gravity \cite{deAraujo:2021zma,Solanki:2021fzo} and references therein). Gravity modification could solve many problems of GR, namely dark energy problem, inflation and late time accelerated expansion. Apart from presenting new geometrodynamical terms in the Einstein-Hilbert (further - EH) action integral, we could also introduce additional matter fields to solve that problems. In the current paper we modified GR Lagrangian by using Bose-Einstein Condensate, Kalb-Ramond and $U(1)$ gauge fields.

\subsection{Bose-Einstein Condensate as Dark Matter / Dark Energy model}
As a first additional matter field of our consideration we have chosen the case with Bose-Einstein condensate. Bose-Einstein condensate is created from the set of Bose gas particles in the same ground state at the very low temperatures near the absolute zero. It is known that Bose-Einstein condensates could be formed from the Cold Dark Matter (CDM) axions by thermalisation process \cite{PhysRevLett.103.111301}. Such BEC dark matter could viably recreate galaxy rotational curves, which was showed in the work of \cite{Craciun2020} on the example of SPARC galactic rotation curves data. Finally, BE condensates could not only act as DM, but also as dark energy, which was shown in the following study \cite{Das_2015}. As well, it was reported by \cite{Fukuyama:2009vzr} that Bose-Einstein condensation phase of the boson field could be present in the early universe, and later that could lead to the formation of dark matter - dark energy unification. It also worth to notice that in the aforementioned model matter density fluctuations are located in the allowed by observational data bounds.

Finally, there was some work done in the field of Bose-Einstein condensate stars, generally adopting hydrodynamical representation for BEC wave function and using polytropic Equation of State. For example, BEC stars were investigated in the works of \cite{Mukherjee:2014kqa,Danila:2015qla,Madarassy:2014jfa}.

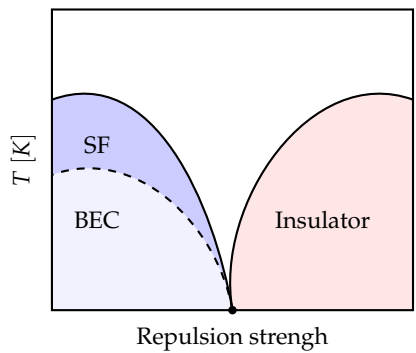
\begin{figure}[!htbp]
    \centering
\begin{tikzpicture}[scale=0.4]
  \message{Phase diagrams 2^^J}
  
  \def\xtick#1#2{\draw[thick] (#1)++(0,.2) --++ (0,-.4) node[below=-.5pt,scale=0.7] {#2};}
  \def\ytick#1#2{\draw[thick] (#1)++(.2,0) --++ (-.4,0) node[left=-.5pt,scale=0.7] {#2};}
  
  % COORDINATES
  \coordinate (Z) at (0,0);
  \coordinate (N1) at (0,7);
  \coordinate (N2) at (11,10);
  \coordinate (N3) at (0,4.5);
  \coordinate (NE) at (12,10);
  \coordinate (NW) at (0,10);
  \coordinate (SE) at (12,0);
  \coordinate (W) at (0,5);
  \coordinate (S) at (6,0);
  \coordinate (C) at (12,7); % critical
  \coordinate (T) at (6,0); % triple
  
  % PATHS
  \def\SL{(T) to[out=100,in=20] (N1)}
  \def\SLL{(T) to[out=100,in=20] (N3)}
  \def\LG{(T) to[out=100,in=160] (C)}
  \def\atm{(0,5.5) -- (12,5.5)}
  \path[name path=SLL] \SLL;
  \path[name path=SL] \SL;
  \path[name path=LG] \LG;
  \path[name path=atm] \atm;
  
  % REGIONS
  \fill[mylightblue] \SL -- (N1) -- (Z) -- cycle;
  \fill[blue!5] \SLL -- (N3) -- (Z) -- cycle;
  \fill[mylightred] \LG -- (C) -- (SE) -- cycle;
  \node at (1.5,3) {BEC};
   \node at (1.5,5.5) {SF};
  \node at (9,3) {Insulator};

  % POINTS
  \fill (T) circle (4pt);

  % LINES
  \draw[thick] \LG;
  \draw[thick] \SL;
  \draw[thick,dashed] \SLL;
  % AXES
  \draw[thick] (Z) rectangle (NE);
  \node[left=3pt,above,rotate=90] at (W) {$T$ $[K]$};
  \node[below=3pt] at (S) {Repulsion strengh};
  
\end{tikzpicture}
    \caption{Phase transition diagram for Boson fluids at small temperature}
    \label{fig:222}
\end{figure}

In the current paper we will use an assumption of zero temperature BEC, namely pure BEC and an assumption of small repulsion strength. Only with these assumptions Gross–Pitaevskii equation could properly describe wave function of the Bose-Einstein condensate (for more details, see phase transition diagram at the Figure (\ref{fig:222})). On the phase transition diagram, there is a couple of phases of Boson fluids present, related to the different values of temperature $T$ and repulsion strength. It could be noticed that fluid with $T\gg0$ and relatively small repulsion strength has superfluid behavior (i.e. vanishing viscosity). However, for smaller values of temperature (assuming that the fluid is in thermodynamical equilibrium), Bose-Einstein Condensate could be formed (pure BEC appears at $T=0$). On the other hand, if we assume big values of repulsion strength, fluid becomes insulator.

\subsection{Kalb-Ramond field and the rise of bouncing cosmology}
Second additional matter field is fully antisymmetric rank-2 tensor field, namely Kalb-Ramond (or B) field. Such fields as Kalb-Ramond one is essential in reproduction of the low energy effective string actions \cite{PhysRevLett.89.121101}. Moreover, massless KR field could occur in the critical and massless string spectrum due to the compactification to four dimensions \cite{Schwarz:2000ew}. Additionally, this field could be the source of spontaneous Lorentz Symmetry Breaking (LSB), if self-interaction potential and non-vanishing vacuum expectation value (VEV) are assumed \cite{PhysRevD.81.065028}. In the pioneering work \cite{Hell_2022} it was shown that KR field could be dual to the massive spin-1 field, namely Proca field. This property could be used in the study of exotic Proca, Boson-Proca stars.

It is also worth to notice that KR field could induce cosmological bounce, which was shown by \cite{nair2021kalbramond} in the modified teleparallel cosmology and that black holes with KR VEV background behave like Reissner-N\" ordstrom black holes, despite the charge is absent \cite{Lessa:2019bgi}.

\subsection{Gauge field background}
Gauge field imposing $U(1)$ local gauge symmetry (namely photon gauge field) is the third and the last exotic matter field that we consider in our study. Stars surrounded by such gauge fields are called gauged boson stars and now these kind of stars and their shells are widely studied in the literature \cite{Kunz2021mbm,Herdeiro:2021jgc,Liu:2020uaz,SalazarLandea:2016bys}. In this articles numerical evidence for the existence of gauged boson stars in asymptotically flat/asymptotically AdS spacetimes was given and it was reported that such solutions similarly to the Kerr-Newmann black holes preserve nonzero electric charge and magnetic dipole momentum \cite{Kichakova2013}. As well, asymptotically AdS boson stars play important role in the AdS/CFT (Conformal Field Theory) correspondence. 

\subsection{Article Organisation}
Our article is organised as follows: in the Section (\ref{sec:1}) we provide a general introduction into the topic of compact (relativistic) objects and their kind, into the gravity modification and problems of GR. As well, we discuss each exotic matter field of our consideration in the separated subsections. On the other hand, in the Section (\ref{sec:22}) we derive field equations for perfect fluid stress-energy tensor in terms of Einsteinian gravity and present the formalism that is being used in our study, introduce viable Finch-Skea metric potentials. Further, in the next Section (\ref{sec:2}) we probe the Finch-Skea compact stars in the presence of pure Bose-Einstein condensate. As well, in the Section (\ref{sec:3}) we investigate our Kalb-Ramond spherically symmetric stellar solutions, in the Section (\ref{sec:4}) case with minimally coupled to gravity gauge field imposing $U(1)$ local gauge symmetry. As a final note, in the Section (\ref{sec:5}) we provide concluding remarks on the main topics of our study.

\section{Finch-Skea stars and Einstein gravity} \label{sec:22}
For the regular GR gravity, Einstein-Hilbert action integral reads:
\begin{equation}
    \mathcal{S}[g,\Gamma,\Psi_i]=\int_\mathcal{M}d^4x\sqrt{-g}\frac{1}{2\kappa^2}(\mathcal{R}+\mathcal{L}(\Psi_i))
    \label{eq:1}
\end{equation}
where $\mathcal{R}$ is common Ricci scalar curvature, $g=\det g_{\mu\nu}=\prod^{3}_{\mu,\nu=0}g_{\mu\nu}$ is the determinant of metric tensor and $\kappa$ is the well-known Einstein gravitational constant. Finally, we define $\mathcal{L}(\Psi_i)$ as a Lagrangian density for additional matter fields $\Psi_i$, $\Gamma$ is torsionless metric-affine connection. Varying the aforementioned EH action with respect to the metric tensor inversion $g^{\mu\nu}$, we could get the set of Einstein Field Equations (further - EFE's):
\begin{equation}
    G_{\mu\nu}=\kappa T_{\mu\nu}
\end{equation}
for the case with flat background spacetime. In the equation above $T_{\mu\nu}$ is defined as stress-energy tensor
\begin{equation}
    T_{\mu\nu} = -\frac{2}{\sqrt{-g}}\frac{\delta(\sqrt{-g}\mathcal{L}(\Psi_i))}{\delta g^{\mu\nu}}
    \label{eq:3}
\end{equation}
In our paper stress-energy-momentum tensor is assumed to be anisotropic and perfect fluid one, and therefore
\begin{equation}
    T_{\mu}^{\nu} =  (\rho+p_t)U_\mu U^\nu - p_t\delta^\nu_\mu + (p_r-p_t)V_\mu V^\nu
\end{equation}
Here, as usual $\rho$ is energy density and $p_r$, $p_t$ are radial, tangential pressures respectively, $U_\mu$ is normalised by $U^\mu U_\mu=-1$ timelike four-velocity and $V_\mu$ is spacelike (radial) four-vector.

Let us consider spherically symmetric line element with metric signature $(+;-;-;-)$:
\begin{equation}
    ds^2 =e^{\nu(r)}dt^2-e^{\lambda(r)}dr^2-r^2d\Omega^2_{D-2}
\end{equation}
where $e^{\nu(r)}$ and $e^{\lambda(r)}$ are metric potentials and $d\Omega^2_{D-2}$ is the $D-2$ dimensional unit sphere line element:
\begin{equation}
    \begin{gathered}
    d\Omega^2_{D-2}=d\theta_1^2+\sin^2\theta_1d\theta_2^2+\sin^2\theta_1\sin^2\theta_2d\theta_3^2+...+\bigg(\prod^{D-3}_{j=1}\sin^2\theta_j\bigg)d\theta^2_{D-2}
    \end{gathered}
\end{equation}
In the further investigation we will restrict our analysis to the $D=(3+1)$ dimensions. For that case, stress-energy-momentum tensor components are \cite{Bhar2017}
\begin{equation}
    \kappa \rho=\frac{1-e^{-\lambda(r)}}{r^2}+\frac{e^{-\lambda(r)}\lambda'(r)}{r}
\end{equation}
\begin{equation}
   \kappa p_r=\frac{e^{-\lambda(r)}-1}{r^2}+\frac{e^{-\lambda(r)}\nu'(r)}{r}
\end{equation}
\begin{equation}
   \kappa p_t = e^{-\lambda}\left(\frac{\nu''(r)}{2}+\frac{\nu'(r)^2}{4}-\frac{\nu'(r)\lambda'(r)}{4}+\frac{\nu'(r)-\lambda'(r)}{2r}\right)
\end{equation}
\subsection{Finch-Skea symmetry}
Throughout the paper we will consider that our compact star solution impose Finch-Skea symmetry and have physically viable, non-singular metric potentials of form \cite{Finch_1989}:
\begin{equation}
    e^{\nu(r)}=\left(A+\frac{1}{2}Br\sqrt{r^2C}\right)^2
\end{equation}
\begin{equation}
    e^{\lambda(r)}=\left(1+Cr^2\right)
\end{equation}
where $A$, $B$ and $C$ have constant values and are called Finch-Skea coefficients. Using Finch-Skea ansatz above, we could rewrite field equations:
\begin{equation}
   \kappa \rho=\frac{C \left(C r^2+3\right)}{\left(C r^2+1\right)^2}
\end{equation}
\begin{equation}
   \kappa p_r=-\frac{C \left(2 A \sqrt{C r^2}+B r \left(C r^2-4\right)\right)}{\left(C r^2+1\right)
   \left(2 A \sqrt{C r^2}+B C r^3\right)}
\end{equation}
\begin{equation}
  \kappa p_t = \frac{C \left(B r \left(C r^2+4\right)-2 A \sqrt{C r^2}\right)}{\left(C r^2+1\right)^2
   \left(2 A \sqrt{C r^2}+B C r^3\right)}
\end{equation}
In the following subsection we are going to derive exact solutions for Finch-Skea coefficients using junction conditions at the boundary.
\subsection{Junction conditions}
Extra matching conditions for spherically symmetric relativistic objects were provided by Goswami et al. \cite{PhysRevD.90.084011}, and it was shown that constraints on the thermodynamic properties and stellar structure are purely mathematical \cite{Pandya2021}. For spacetime with axymptotically flat background, exterior region could be considered as a Schwarzschild’s spacetime:
\begin{equation}
    ds^2 = \bigg(1-\frac{2M}{r}\bigg)dt^2 - \bigg(1-\frac{2M}{r}\bigg)^{-1}dr^2 - r^2 d\theta^2 - r^2 \sin^2 \theta d\phi^2
\end{equation}
where $M$ indicates total gravastar mass. The above line element imply conditions of metric potential continuity at the boundary:
\begin{equation}
    \mathrm{Continuity\;of\;}g_{tt}:\quad 1-\frac{2M}{R}=e^{\nu(r)}
\end{equation}
\begin{equation}
    \mathrm{Continuity\;of\;}\frac{\partial g_{tt}}{\partial r}:\quad \frac{2M}{R^2}=-B \left(2 A \sqrt{C R^2}+B C R^3\right)
\end{equation}
\begin{equation}
    \mathrm{Continuity\;of\;}g_{rr}:\quad \bigg(1-\frac{2M}{R}\bigg)^{-1}=e^{\lambda(r)}
\end{equation}
\begin{equation}
    \mathrm{Vacuum\;condition}:\quad p\rvert_{r=R}=0
\end{equation}
The solutions for above continuity conditions are:
\begin{equation}
    A=\frac{3 M-2 R}{2 \sqrt{R} \sqrt{R-2 M}}
\end{equation}
\begin{equation}
    B=\frac{\sqrt{\frac{M}{R-2 M}} \sqrt{R-2 M}}{\sqrt{2} R^{3/2}}
\end{equation}
\begin{equation}
    C=\frac{2 M}{R^2 (R-2 M)}
\end{equation}
\begin{figure}[!htbp]
    \centering
    \includegraphics[width=\textwidth]{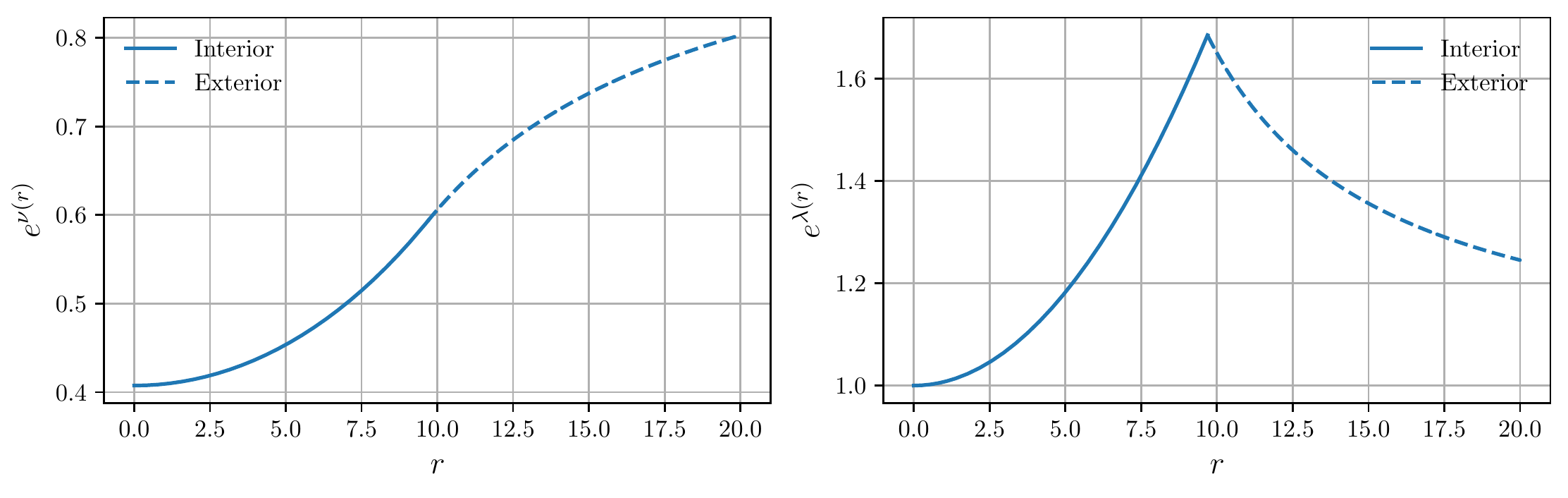}
    \caption{Finch-Skea interior and exterior metric potentials with adopted junction conditions for compact star PSRJ1416-2230}
    \label{fig:6555}
\end{figure}

We as well plotted metric potentials using junction conditions defined above on the Figure (\ref{fig:6555}) for relativistic star PSRJ1416-2230, which has mass $M+1.97$ and Radius $R=9.69$. Now we could proceed further and define Bose-Einstein condensate (BEC) field minimally coupled to gravity.

\section{Minimally coupled Bose-Einstein condensate} \label{sec:2}
\begin{figure}[!htbp]
    \centering
    \includegraphics[width=\textwidth]{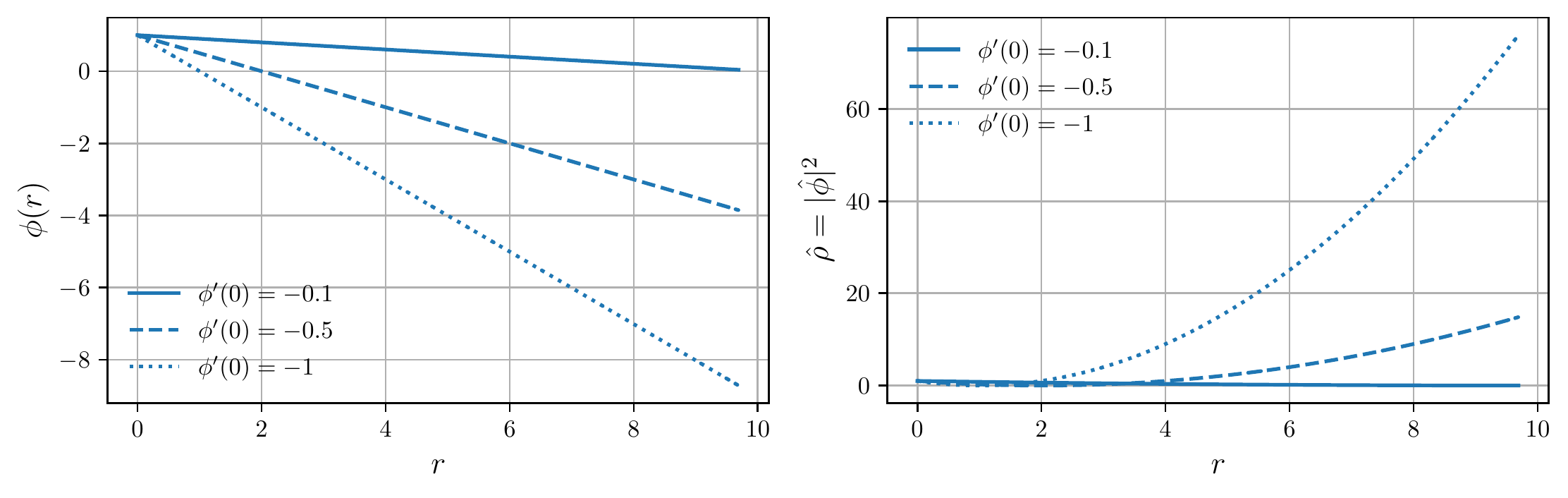}
    \caption{(\textit{left plot}) Solutions for scalar field $\phi(r)$ governed by Gross–Pitaevskii equation, (\textit{right plot}) probability density for Bose-Einstein condensate. To obtain solutions, we assumed that $m=\eta=10^{-8}$, $\omega=10^{-2}$. As well, mass and radius were similar to compact star PSRJ1416-2230 ($M=1.97$ and $R=9.69$)}
    \label{fig:1}
\end{figure}

Generally, the ground state of Bose-Einstein condensate could be described by the complex scalar field \cite{Fagnocchi:2010sn}. If we assume Minimally-coupled to Einsteinian gravity Bose-Einstein field, classical GR Lagrangian (\ref{eq:1}) transforms as
\begin{equation}
        \mathcal{S}[g,\Gamma,\Psi_i,\hat{\phi}]=\int_\mathcal{M}d^4x\sqrt{-g}\frac{1}{2\kappa^2}(\mathcal{R}+\mathcal{L}(\Psi_i)+\mathcal{L}(\hat{\phi}))
    \label{eq:23}
\end{equation}
where we define \cite{Bettoni_2014}
\begin{equation}
    \mathcal{L}(\hat{\phi})=-g_{\mu\nu}\nabla^\mu\hat{\phi}^\dagger \nabla^\nu \hat{\phi}-m^2\hat{\rho}-U(\hat{\rho})
\end{equation}
Here, obviously $\hat{\phi}$ is Bose-Einstein field, $\hat{\phi}^\dagger$ is its complex conjugate, $m^2$ is scalar Bose field mass and finally $U(\hat{\rho})$ is the so-called self-interaction potential, which could be expanded in a series with many-body interaction terms (external potential $V(\hat{\rho})$ for Bose field is absent) \cite{PhysRevD.9.3320,PhysRevD.9.3357}:
\begin{equation}
    U(\hat{\rho})=\frac{\lambda_2}{2}\hat{\rho}^2+\frac{\lambda_3}{6}\hat{\rho}^3+...-\frac{\lambda_2}{8}T^2\hat{\rho}+\frac{\pi^2}{90}T^4\hat{\rho}^2
\end{equation}
where first term corresponds to the regular two particle interaction, $\rho=\hat{\phi}^\dagger\hat{\phi}$ is the Bose field probability density. For the sake of simplicity, in the definition of self-interaction potential we will use only the terms up to quadratic, and therefore we redefine set of self-interaction coupling coefficients to $\eta$. As well, we assume that our Bose-Einstein condensate is pure, and therefore $T=0$. Using (\ref{eq:3}) we could easily obtain stress-energy tensor for our Bose field:
\begin{equation}
    \begin{gathered}
    T_{\mu\nu}^{\hat{\phi}}=\nabla_\mu \hat{\phi}^\dagger \nabla_\nu \hat{\phi}+\nabla_\mu \hat{\phi}\nabla_\nu \hat{\phi}^\dagger -g_{\mu\nu}\bigg(g^{\alpha\beta}\nabla_\alpha \hat{\phi}^\dagger \nabla_\beta \hat{\phi} +m^2\hat{\rho}+U(\hat{\rho})\bigg)
    \end{gathered}
\end{equation}
Since we are working with scalar field, for the first order covariant derivatives further we imply proper transformation $\nabla_\mu \hat{\phi}\to \partial_\mu \hat{\phi}$. To numerically derive Bose field it is also useful to introduce modified massive Klein-Gordon equation:
\begin{equation}
    g_{\mu\nu}\nabla^\mu \nabla^\nu \hat{\phi}-\bigg(m^2+U'(\hat{\rho})\bigg)\hat{\phi}=0
\end{equation}
Using the assumption of harmonic time dependence we could impose the transformation
\begin{equation}
    \hat{\phi}=\exp(-i\omega t)\phi(r)
\end{equation}
where $\phi(r)$ is the real function that depends only on radial coordinate $r$, and therefore $\hat{\rho}=\hat{\phi}\hat{\phi}^\dagger=\phi^2$. Using the aforementioned assumption. Klein-Gordon equation reduces to Gross–Pitaevskii–like equation \cite{Matos:2016ryp}:
\begin{equation}
    i\hslash \nabla_t \hat{\phi}=\bigg(-\frac{\hslash^2}{2m}\nabla^i\nabla_i+\underbrace{mV_{\mathrm{ext}}}_\text{0}+\eta\hat{\rho}^2\bigg)\hat{\phi}
\end{equation}
where 
\begin{equation}
    \nabla_i\hat{\phi} = \frac{1}{\sqrt{-g}}\partial_i(\sqrt{-g}g^{ij}\partial_i\hat{\phi}) 
\end{equation}
and $V_{\mathrm{ext}}$ is external integral. We will solve the equation above numerically for the set of initial conditions $\phi(0)= 10$ and $\phi'(0)=C$. Solutions for above equation have viable behavior for $\omega\gg|\eta|\land m$ and for relatively small values of this parameters. We plot the oscillating solution for $\phi(r)$ on the Figure (\ref{fig:1}). 

Finally, we could derive field equations from modified Einstein-Hilbert action (\ref{eq:23}):
\begin{equation}
    \kappa (\rho+\rho^{\hat{\phi}})=\frac{1-e^{-\lambda(r)}}{r^2}+\frac{e^{-\lambda(r)}\lambda'(r)}{r}
\end{equation}
\begin{equation}
   \kappa (p_r+p_r^{\hat{\phi}})=\frac{e^{-\lambda(r)}-1}{r^2}+\frac{e^{-\lambda(r)}\nu'(r)}{r}
\end{equation}
\begin{equation}
   \kappa (p_t+p_t^{\hat{\phi}}) = e^{-\lambda}\left(\frac{\nu''(r)}{2}+\frac{\nu'(r)^2}{4}-\frac{\nu'(r)\lambda'(r)}{4}+\frac{\nu'(r)-\lambda'(r)}{2r}\right)
\end{equation}
where \cite{doi:10.1142/S0218271821300068}
\begin{equation}
	\rho^{\hat{\phi}} = -e^{-\nu}\omega^2\phi^2 - e^{-\lambda}(\phi')^2 + V(\phi) 
\end{equation}
\begin{equation}
	p_r^{\hat{\phi}} = -e^{-\nu}\omega^2\phi^2 - e^{-\lambda}(\phi')^2 - V(\phi) 
\end{equation}
\begin{equation}
	p_t^{\hat{\phi}} = -e^{-\nu}\omega^2\phi^2 + e^{-\lambda}(\phi')^2 - V(\phi) 
\end{equation}
Here, we define $V(\Phi)=m^2\hat{\rho}^2/2+U(\hat{\rho})$.
In the next subsection we are going to probe the behavior of Finch-Skea star minimally coupled to Bose-Einstein condensate.
\subsection{Energy Conditions}
Energy conditions are probes of relativistic model viability. Generally, there exists four energy conditions, namely Weak Energy Condition (WEC), Null Energy Condition
(NEC), Strong Energy Condition (SEC) and Dominant Energy Condition (DEC). For model to be physically plausible, all of the aforementioned energy conditions must be satisfied at the every point of manifold. EC's, derived from temporal Raychaudhuri equations are defined as follows:
\begin{itemize}
\item Null Energy Condition (NEC): $\rho +p_{r}\geq 0$ and $\rho +p_{t}\geq 0
$

\item Weak Energy Condition (WEC) $\rho >0$ and $\rho +p_{r}\geq 0$ and $%
\rho +p_{t}\geq 0$

\item Dominant Energy Condition (DEC): $\rho -|p_{r}|\geq 0$ and $\rho
-|p_{t}|\geq 0$

\item Strong Energy Condition (SEC): $\rho +p_{r}+2p_{t}\geq 0$
\end{itemize}
\begin{figure}[!htbp]
    \centering
    \includegraphics[width=\textwidth]{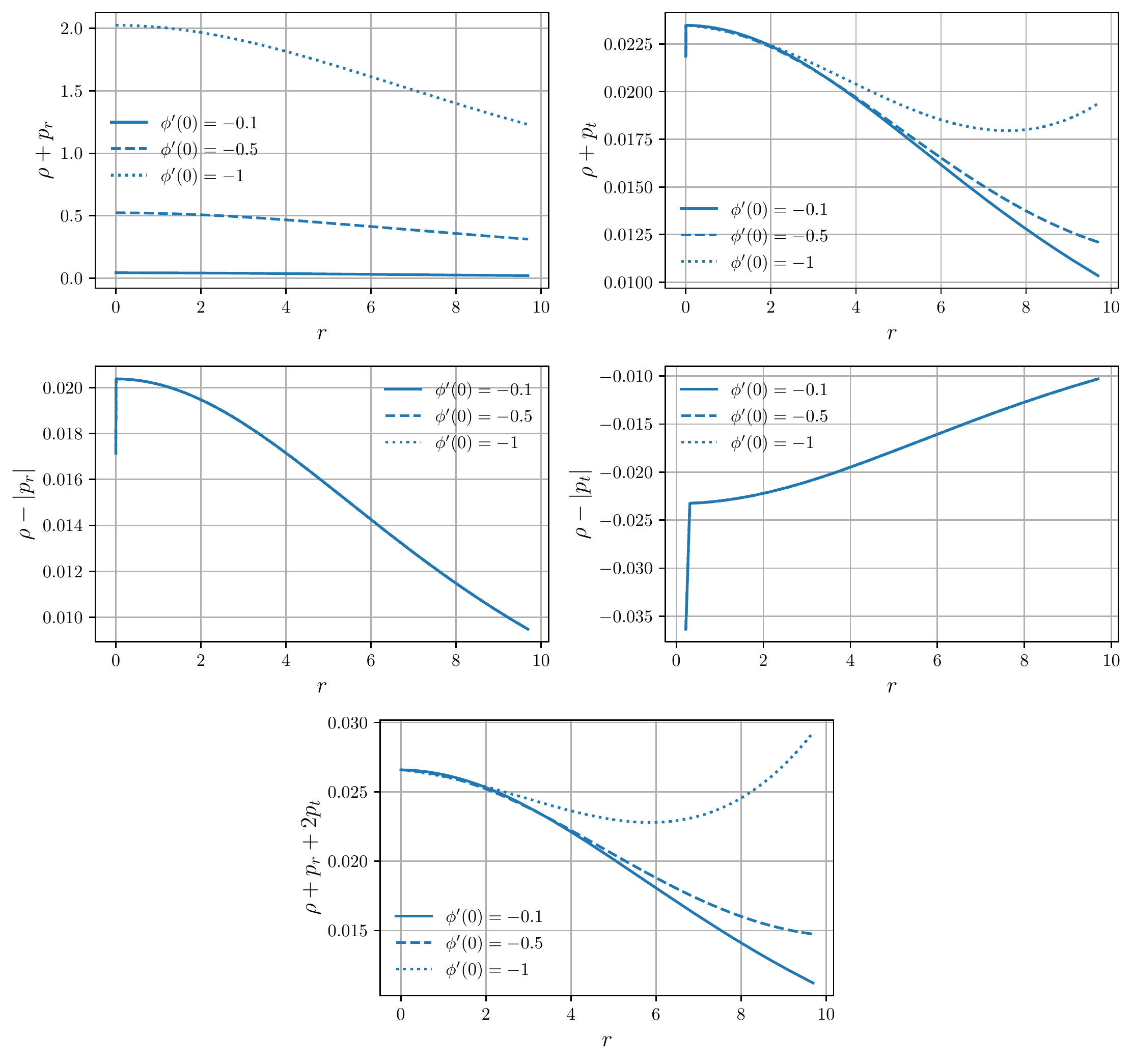}
    \caption{Null, Dominant and Strong energy conditions for Finch-Skea PSRJ1416-2230 star minimally coupled to Bose-Einstein condensate (we assumed that $m=\eta=10^{-8}$, $\omega=10^{-2}$)}
    \label{fig:3}
\end{figure}

We plot energy conditions for Finch-Skea PSRJ1416-2230 star minimally coupled to Bose-Einstein condensate on the Figure (\ref{fig:3}). As one may obviously notice, NEC is validated for tangential and radial pressures, DEC is validated for both pressure kinds, SEC is obeyed everywhere. It is worth to state that Null, Dominant and Strong Energy Conditions will be still satisfied for every relatively small and positive values of Bose field mass squared, frequency and self-coupling constant $\eta$.  

\subsection{Gradients and Equation of State}
\begin{figure}[!htbp]
    \centering
    \includegraphics[width=\textwidth]{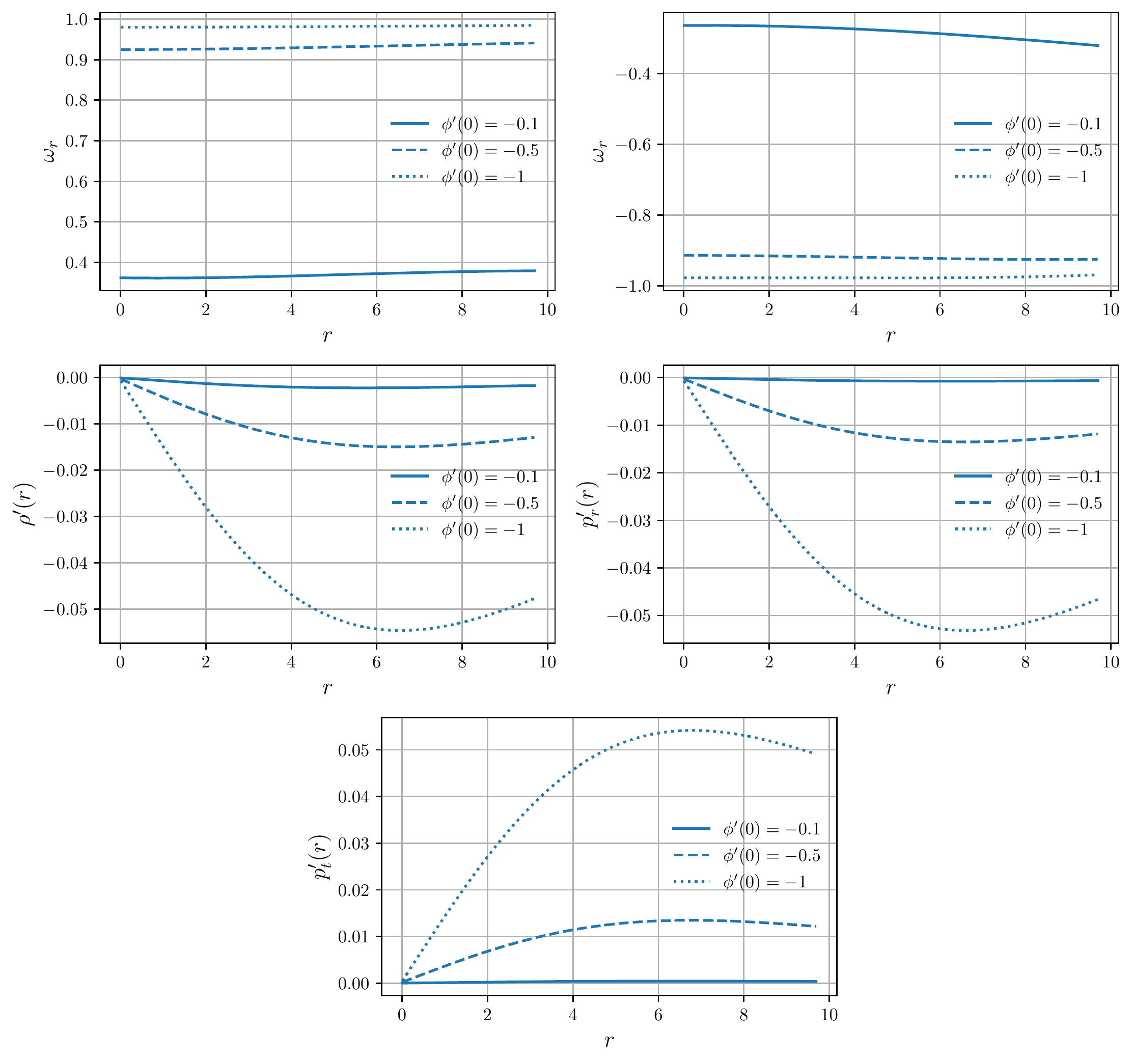}
    \caption{(\textit{left plot}) Equation of State for radial and tangential pressures (parameter values are the same as for EC's), (\textit{right plot}) Gradient of the energy density and anisotropic pressures}
    \label{fig:4}
\end{figure}

To probe the nature of matter content in the Finch-Skea star interior region we will use the Eqution of State (further - EoS) parameter, which is defined as
\begin{equation}
    \omega_r=p_r/\rho,\quad \omega_t=p_t/\rho
\end{equation}
As well, we could depict the energy density, anisotropic pressure gradients by simply calculating $\rho'(r)$ and $p_r'(r)$, $p_t'(r)$. We illustrated EoS and stress-energy tensor component gradients on the Figure (\ref{fig:4}). As we see, at the stellar origin radial EoS is asymptotically ($\phi'(0)\to-\infty$) described by stiff Zeldovich fluid and tangential one by dark-energy fluid, but for all negative values of $\phi'(0)$ initial condition radial fluid is regular one in the bounds $\omega\in(0,1)$ and tangential is quintessence in the bounds $\omega\in(-1,0)$. On the other hand, gradient of energy density, anisotropic pressures could have both positive and negative values in the stellar interior, but generally because of the asymptotical flatness, gradients vanish at the region $r\to \infty$.

\subsection{Tolman-Oppenheimer-Volkoff stability}
\begin{figure}[!htbp]
    \centering
    \includegraphics[width=\textwidth]{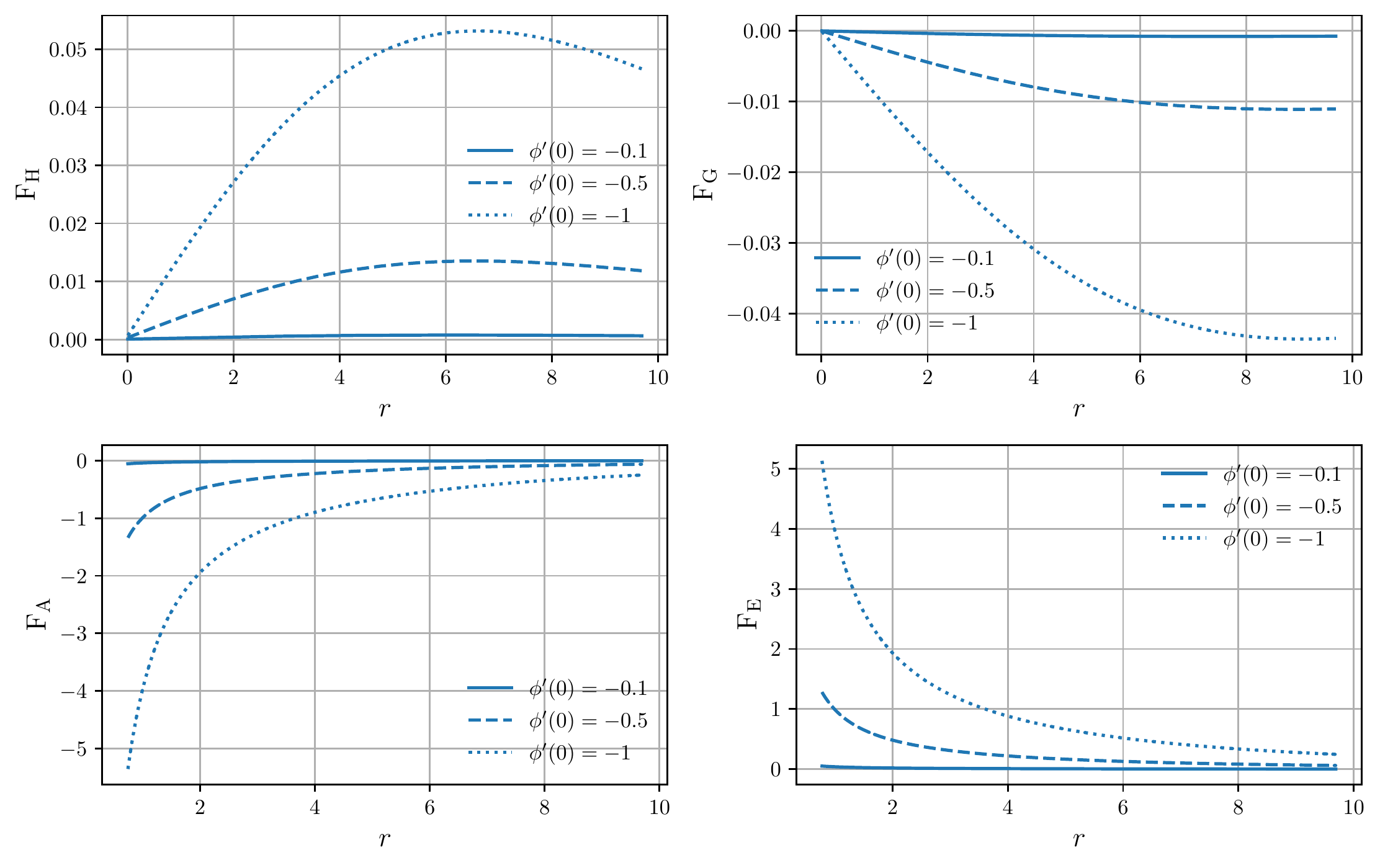}
    \caption{Modified TOV forces for Finch-Skea star minimally coupled to Bose-Einstein condensate}
    \label{fig:5}
\end{figure}

Stability of the matter content in the stellar interior could be investigated using the well known Tolman-Oppenheimer-Volkoff equilibrium condition, which is given below in its modified form \cite{Oppenheimer1939,Poncede1993,Rahaman2014,Tolman1939}:
\begin{equation}
-\frac{dp_{r}}{dr}-\frac{\nu^{'}(r)}{2}(\rho+p_{r})+\frac{2}{r}(p_{t}-p_{r})+F_{\mathrm{ex}}=0
\label{eq:34}
\end{equation}
As one may obviously notice, in the modified form of TOV equilibrium condition present one extra force, namely $F_{\mathrm{ex}}$, which is present because of the stress-energy-momentum tensor discontinuity ($\nabla^\mu T_{\mu\nu}\neq0$) to hold relativistic object stable. As well, in classical and modified TOV's present three additional forces: hydrodynamical $F_\mathrm{H}$, gravitational $F_{\mathrm{G}}$ and anisotropic $F_{\mathrm{A}}$:
\begin{equation}
F_H=-\frac{dp_{r}}{dr},\;\;\;\;\;\;\;\;F_A=\frac{2}{r}(p_{t}-p_{r}), \;\;\;\;\;\;\;\;F_G=-\frac{\nu^{'}}{2}(\rho+p_{r}),
\end{equation}
and therefore, we could easily rewrite MTOV (\ref{eq:34}):
\begin{equation}\label{t}
F_A+F_G+F_H+F_{ex}=0
\end{equation}
We plot the solutions for each force present in MTOV on the Figure (\ref{fig:5}). It is noticable that both hydrodynamical force $F_H$ and extra force $F_{ex}$ have positive values at the whole interior radial domain, gravitational and anisotropic forces are always negative and asymptotically vanish. Hence, for Finch-Skea star in the presence of Bose-Einstein condensate modified TOV equilibrium is satisfied (relativistic star is stable) if extra force is present.

\subsection{Adiabatic index}
Adiabatic index is the tool that could be used to probe relativistic object via adiabatic perturbations. Firstly, it was introduced in the work of Chandrasekhar \cite{Chandrasekhar1964}. Chandrasekhar predicted that for the relativistic system to be stable the adiabatic index should exceed $4/3$. This adiabatic index is defined as \cite{Maurya2017}:
\begin{equation}
    \Gamma = \frac{p_r+\rho}{p_r}\frac{dp_r}{d\rho}
    \label{eq:31}
\end{equation}
\begin{figure}[!htbp]
    \centering
    \includegraphics[width=\textwidth]{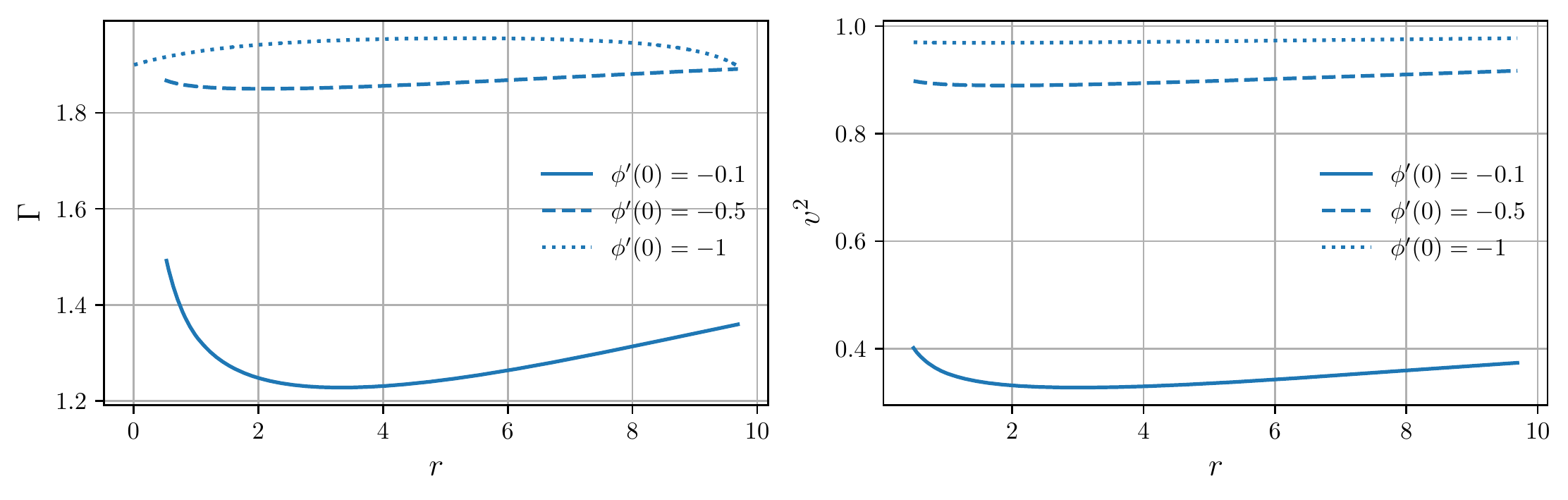}
    \caption{(\textit{first row}) Adiabatic index for Finch-Skea star minimally coupled to Bose-Einstein condensate, (\textit{second row}) speed of sound for Finch-Skea star minimally coupled to Bose-Einstein condensate}
    \label{fig:76}
\end{figure}
Adiabatic index solutions are plotted on the Figure (\ref{fig:76}). Generally, for the same values of parameters, as it was used in the energy conditions subsection, $\Gamma$ constraint satisfied at the area near origin and envelope for small $\phi'(0)<0$ and satisfied everywhere for $\phi'(0)\ll0$. Moreover, stability holds for every relatively small and positive number of $\omega$, $\eta$ and $m$.

\subsection{Surface redshift}
\begin{figure}[!htbp]
    \centering
    \includegraphics[width=0.6\textwidth]{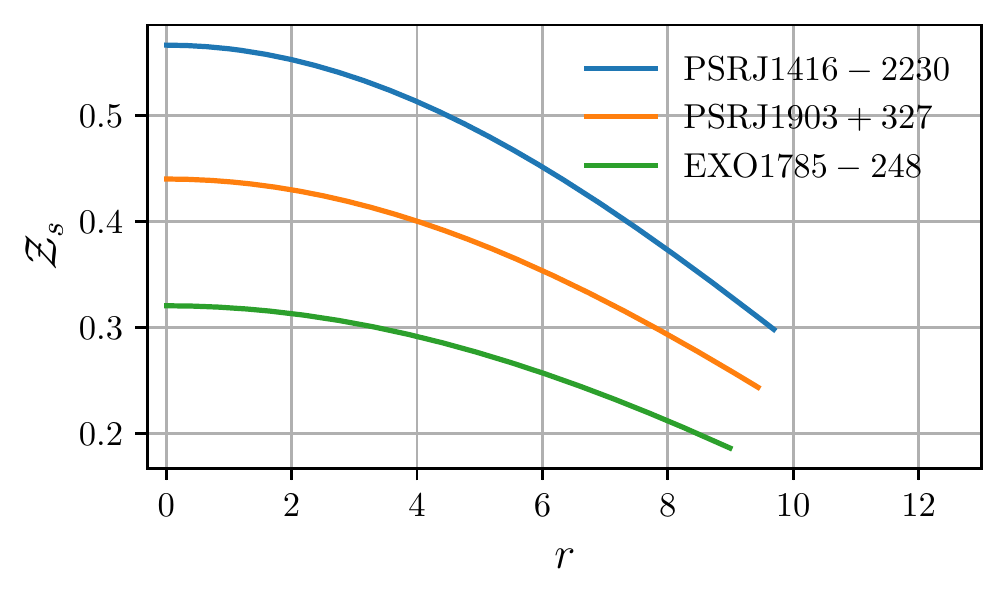}
    \caption{Surface (gravitational) redshift for Finch-Skea stellar interior (parameter values are the same as for EC's)}
    \label{fig:6}
\end{figure}

Finally, we could also define surface redshift:
\begin{equation}
    \mathcal{Z}_s = |g_{tt}|^{-1/2}-1
    \label{eq:34}
\end{equation}
Which for anisotropic matter distribution must not exceed the value of 2. Routinely, we plotted numerical solution for surface redshift for different compact stars on the Figure (\ref{fig:6}). Judging by the plotted data we could conclude that our solution satisfy surface gravitational redshift constraint.

\subsection{Sound of speed}
Another important criterion of matter content viability is the so-called speed of sound. This quantity is defined in the following way (\textbf{derived from the ultra-relativistic hydrodynamics}):
\begin{equation}
    v^2 = \frac{dp_r}{d\rho}\leq c^2 =1
\end{equation}
\textbf{Inequality above needs to be satisfied in order to respect the causality constraints.} Solution for speed of sound is plotted on the third plot of Figure (\ref{fig:76}). As we see, inequality $v^2\leq c^2$ always holds, which is necessary condition. Also, it is important to state that asymptotically $\phi'(0)\to-\infty$ our fluid becomes massless, since $v^2=c^2$.
Since we already probed all of the necessary parameters for minimally coupled Bose-Einstein condensate, we could proceed to the minimally coupled Kalb-Ramond field case.

\section{Minimally coupled Massless Kalb-Ramond field} \label{sec:3}
In this section, we are going to investigate the compact stars admitting Finch-Skea symmetry in the presence of background Kalb-Ramond (further - KB) field. For that particular case, total Einstein-Hilbert action integral is given below:
\begin{equation}
        \mathcal{S}[g,\Gamma,\Psi_i,\hat{\phi},B_{\mu\nu}]=\int_\mathcal{M}d^4x\sqrt{-g}\frac{1}{2\kappa^2}(\mathcal{R}+\mathcal{L}(\Psi_i)))+\int_\mathcal{M}d^4x\sqrt{-g} H_{\mu\nu\alpha}H^{\mu\nu\alpha}
    \label{eq:23}
\end{equation}
In the action above, $H_{\mu\nu\alpha}$ is defined as a Kalb-Ramond field strength:
\begin{equation}
    H_{\mu\nu\alpha}=\partial _\mu B_{\nu\alpha}+\partial _\nu B_{\alpha\mu}+\partial _\alpha B_{\mu\nu}
\end{equation}
where obviously $B_{\mu\nu}$ is the antisymmetric rank 2 tensor, namely Kalb-Ramond field. For KB field we have a set of two EoMs, which are given as follows \cite{PhysRevD.77.044030,PhysRevD.52.5877,2018EPJC...78..531D}:
\begin{equation}
     \frac{1}{\sqrt{-g}}\partial_\mu (\sqrt{-g}H^{\mu\nu\alpha})=0
    \label{eq:41}
\end{equation}
\begin{equation}
    \epsilon^{\mu\nu\alpha\beta}\partial_\mu (\sqrt{-g}H_{\nu\alpha\beta})
\end{equation}
Here, $\epsilon^{\mu\nu\alpha\beta}$ is the well known Levi-Cevita symbol, which is for Minkowskian spacetime equal to unity if it's indices are an even permutation of $(1234)$, equal to $-1$ if indices are odd permutation of $(1234)$ and vanish if one of the indices repeat. For KB field, stress-energy-momentum tensor reads \cite{maluf2022bianchi}:
\begin{equation}
    T^{\mu\nu}_B=\frac{1}{2}H^{\alpha\beta\mu}H^{\nu}_{\;\alpha\beta}-\frac{1}{2}g^{\mu\nu}H^{\alpha\beta\gamma}H_{\alpha\beta\gamma}
\end{equation}
where we consider that there is no external potential present. Solving field equation (\ref{eq:41}), we could obtain the solution for Kalb-Ramond field strength \cite{nair2021kalbramond}:
\begin{equation}
    H_{\mu\nu\alpha}=\epsilon^{\mu\nu\alpha\beta}\partial_\beta \phi
\end{equation}
Here, $\phi$ is the real scalar field, whose evolution is governed by the following equation:
\begin{equation}
    \nabla_\mu\nabla^\mu\phi =0 
\end{equation}
Which is exactly massless wave equation. As usual, we will assume only radial coordinate dependence for real scalar field and as well use set of initial conditions to solve wave equation above numerically: $\phi(0)=0.1$ and $\phi'(0)=C$ (if we assume that radial derivative of scalar field vanish at the origin, scalar field will have constant solution). To numerically plot the results in the presence of KB field we will vary the values of scalar field first order radial derivative at the origin $\phi'(0)$. Finally, using previously defined stress-energy-momentum tensor for KB field we could write down new set of (modified) Einstein Field Equations:
\begin{equation}
    \begin{gathered}
    \rho=e^{\lambda  (r)-4 \nu (r)} \phi  '(r)^2-\frac{e^{-\lambda  (r)} \left(r \lambda 
   '(r)+e^{\lambda  (r)}-1\right)}{r^2}
    \end{gathered}
\end{equation}
\begin{equation}
    \begin{gathered}
    p_r =-\frac{e^{-\lambda  (r)} \left(r \nu '(r)+1\right)}{r^2}+\frac{1}{r^2}+3 e^{-3 \lambda  (r)} 
   \phi '(r)^2
    \end{gathered}
\end{equation}
\begin{equation}
    \begin{gathered}
    p_t =\frac{e^{-\lambda  (r)} \left(\left(r \nu '(r)+2\right) \left(\lambda  '(r)-\nu '(r)\right)-2 r
   \nu ''(r)\right)}{4 r}-\frac{e^{\lambda  (r)} \phi  '(r)^2}{r^8}
    \end{gathered}
\end{equation}
Since our spacetime admits spherical symmetry, without the loss of generality we could work only in the equatorial plane adopting $\theta=\pi/2$.

\subsection{Energy conditions}
\begin{figure}[!htbp]
    \centering
    \includegraphics[width=\textwidth]{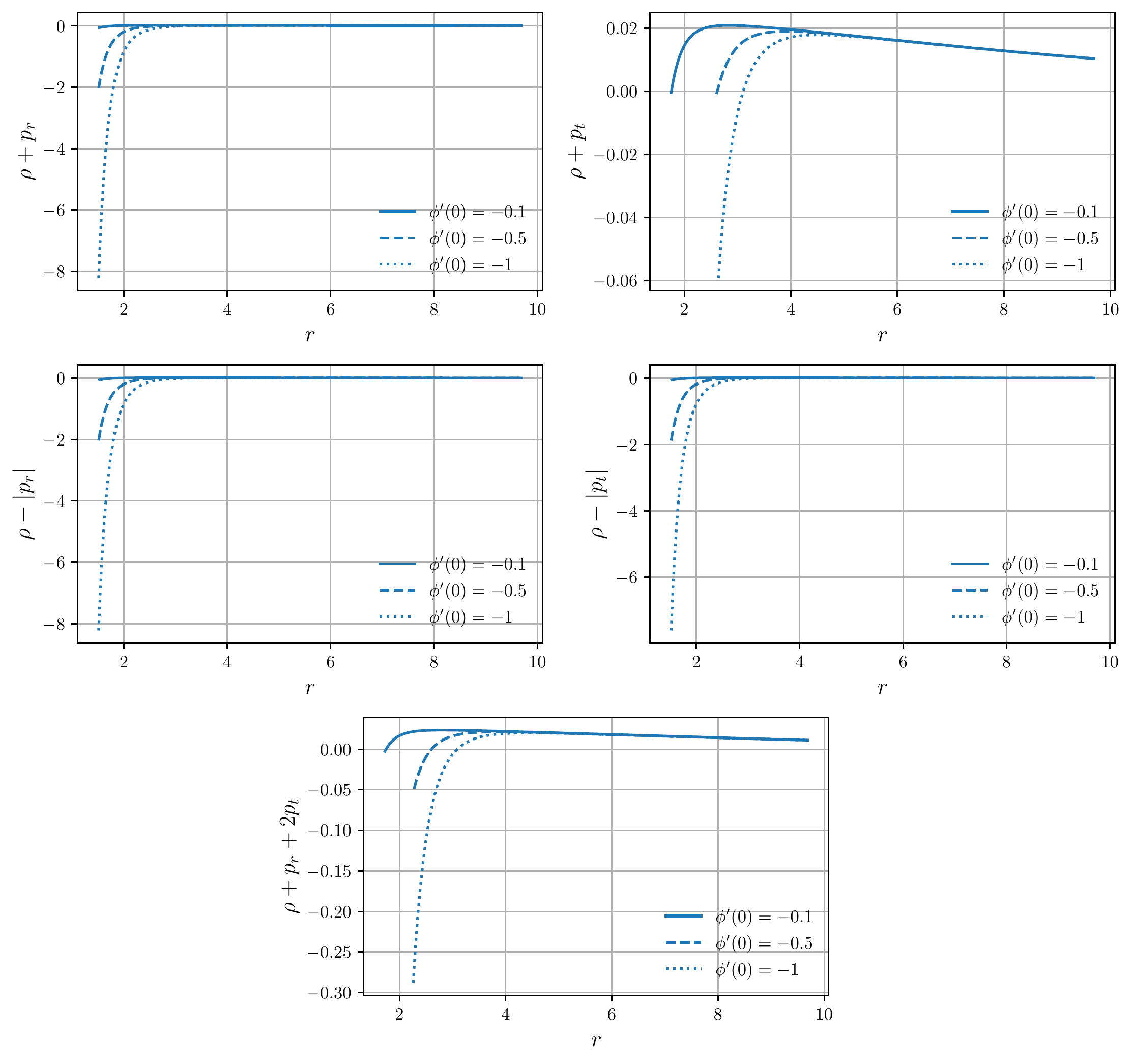}
    \caption{Null, Dominant and Strong energy conditions for Finch-Skea PSRJ1416-2230 star minimally coupled to Kalb-Ramond field}
    \label{fig:8}
\end{figure}

Energy conditions for Finch-Skea star minimally coupled to Kalb-Ramond field are illustrated on the Figure (\ref{fig:8}). Obviously, from that data we could conclude that all of the energy conditions are violated. This conditions could be satisfied if we will take another values of free parameters $\omega$, initial conditions $\phi(0)$ and $\phi'(0)$, but for that case speed of sound will exceed $c^2$, which is forbidden.

\subsection{Gradients and Equation of State}
\begin{figure}[!htbp]
    \centering
    \includegraphics[width=\textwidth]{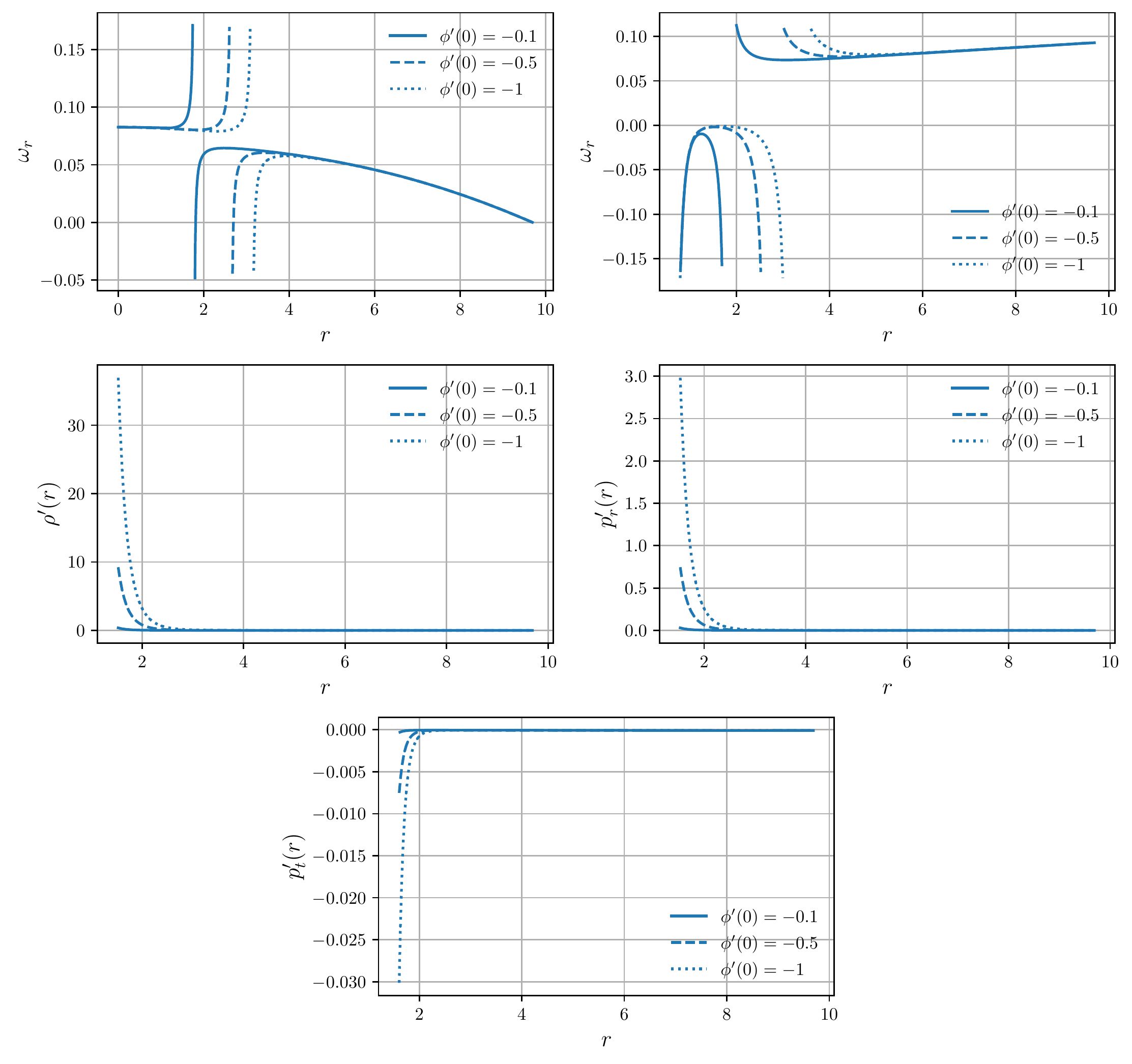}
    \caption{Equation of state and radial derivative of stress-energy tensor components for Finch-Skea star coupled to KB field}
    \label{fig:9}
\end{figure}

In addition to the energy conditions, we as well probed the equation of state for our stellar matter content and energy density, anisotropic pressure radial derivatives behavior. Results are consequently plotted on the Figure (\ref{fig:9}). EoS has regular behavior ($0<\omega<1$) for radial pressure and is negative (quintessence) for tangential pressure (near the stellar core), regular near envelope. Finally, gradients generally are positive for energy density and radial pressure, negative for tangential one.

\subsection{TOV stability}
\begin{figure}[!htbp]
    \centering
    \includegraphics[width=\textwidth]{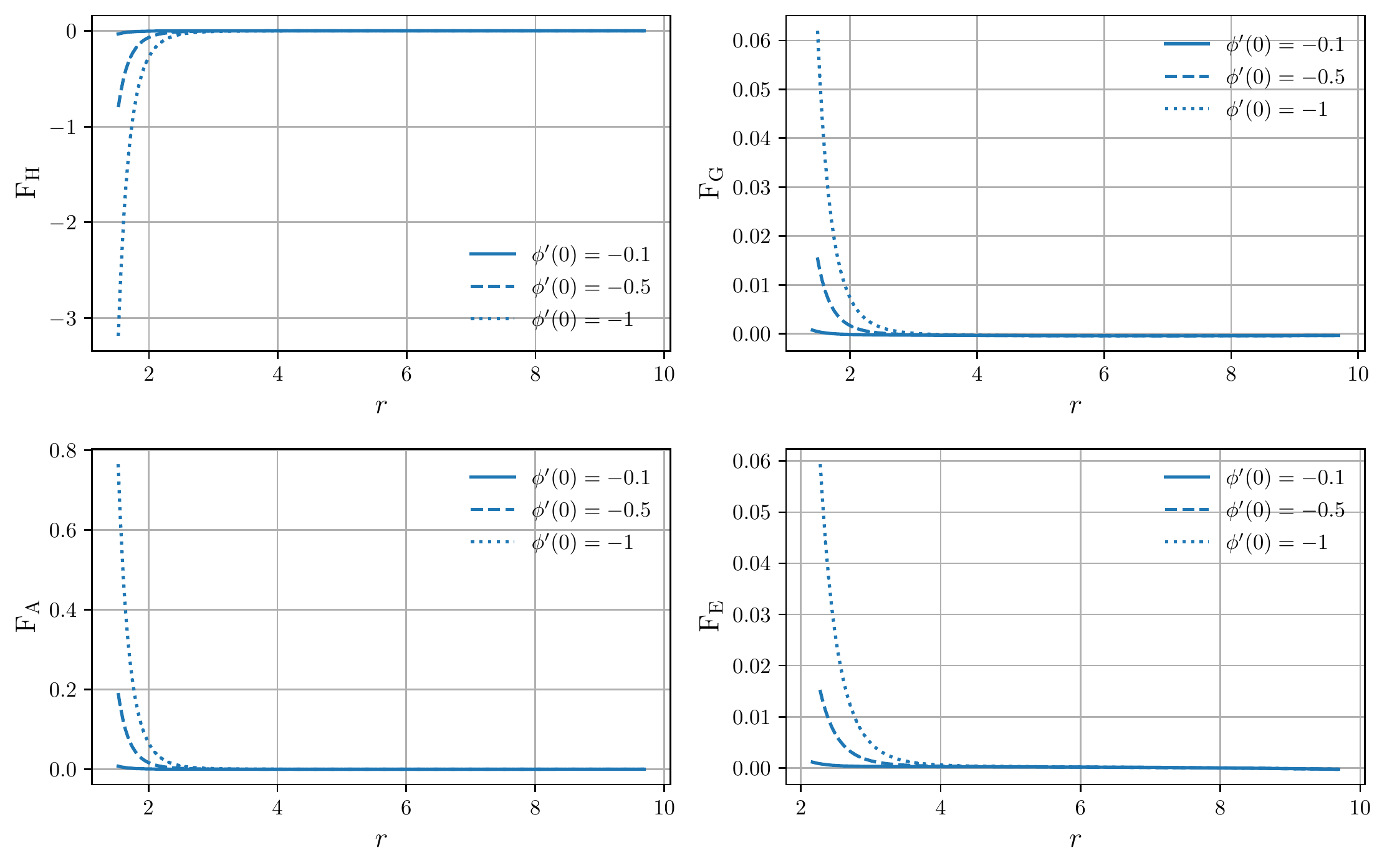}
    \caption{Forces present in the modified TOV equilibrium for Finch-Skea star coupled to KB field}
    \label{fig:10}
\end{figure}

It is also convenient to investigate the TOV stability of our Finch-Skea stellar solution coupled to KB field. Graphical results of such investigation are located on the four plots of Figure (\ref{fig:10}). It is worth to notice that only hydrodynamical force have negative values, gravitational, anisotropic and extra forces therefore have positive values.

\subsection{Adiabatic index}
\begin{figure}[!htbp]
    \centering
    \includegraphics[width=\textwidth]{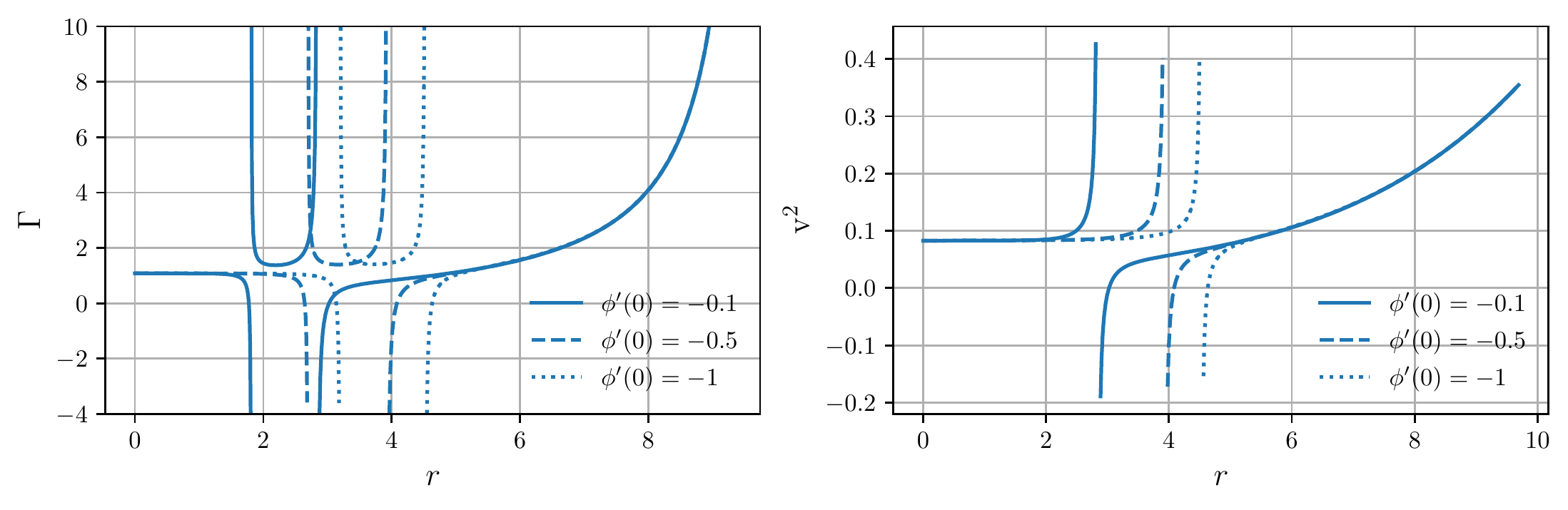}
    \caption{($first\;row$) Adiabatic index for anisotropic Finch-Skea star coupled to KB field, ($second\;row$) sound of speed squared for anisotropic Finch-Skea star coupled to KB field}
    \label{fig:11}
\end{figure}

As a final note, in this subsection we are going to probe the behavior of adiabatic index for our stellar solutions. Results of numerical investigation are properly plotted on the Figure (\ref{fig:11}). One may notice that our stellar solution is stable everywhere except the regions where adiabatic index numerical solution diverge.

\subsection{Speed of sound}
As the final note on minimal KB field in this subsection we are going to investigate the speed of sound for perfect fluid matter content inside Finch-Skea compact star. Results of such investigation are properly plotted on the last, third plot of Figure (\ref{fig:11}). As we see, judging by the data from the plot our fluid is relativistic, but speed of sound does not exceed the speed of light, which is required for fluid to be viable. As well, it is worth to notice that the values of $v^2$ are generally smaller in the presence of KB field (in relation to the classical GR).

\section{$U(1)$ Gauge background field cosmology}\label{sec:4}
In this section we present the final model of our consideration, namely gauged boson Finch-Skea star. This theory has a a gauge field $A_\mu$ admitting $U(1)$ unitary symmetry and massless complex scalar field $\Phi$ with conical potential $V(|\Phi|)=\lambda |\Phi|$. For that kind of model, Lagrangian looks exactly like:
\begin{equation}
        \mathcal{S}[g,\Gamma,\Psi_i,\hat{\phi},A_\mu,\Phi]=\int_\mathcal{M}d^4x\sqrt{-g}\frac{1}{2\kappa^2}(\mathcal{R}+\mathcal{L}(\Psi_i)))-\frac{1}{4}\int_\mathcal{M}d^4x\sqrt{-g}F^{\mu\nu}F_{\mu\nu} -\int_\mathcal{M}d^4x\sqrt{-g}(\mathcal{D}_\mu \Phi)^*(\mathcal{D}^\mu \Phi)-V(|\Phi|)
    \label{eq:49}
\end{equation}
where,
\begin{equation}
   \mathcal{D}_\mu \Phi = \partial_\mu \Phi+ieA_\mu \Phi
\end{equation}
\begin{equation}
    F_{\mu\nu}=\partial_\mu A_\nu - \partial_\nu A_\mu
\end{equation}
Asterisk about expression with gauge derivative denotes usual complex conjugation. For electromagnetic field $F_{\mu\nu}$ and massless scalar field $\Phi$ equations of motion could be easily derived from the minimal action:
\begin{equation}
    \mathcal{D}_\mu(\sqrt{-g}F^{\mu\nu})=-ie\sqrt{-g}[\Phi^*(\mathcal{D}^\nu\Phi)-\Phi(\mathcal{D}^\nu\Phi)^*]
\end{equation}
\begin{equation}
    \mathcal{D}_\mu(\sqrt{-g}\mathcal{D}^\mu\Phi)=\frac{\lambda}{2}\sqrt{-g}\frac{\Phi}{|\Phi|}
\end{equation}
\begin{equation}
   [ \mathcal{D}_\mu(\sqrt{-g}\mathcal{D}^\mu\Phi)]^*=\frac{\lambda}{2}\sqrt{-g}\frac{\Phi^*}{|\Phi|}
\end{equation}
Finally, it is also useful to define stress-energy-momentum tensor for gauge and complex scalar fields respectively
\cite{PhysRevD.93.044014}:
\begin{equation}
    T_{\mu\nu}^F=E_{\mu\nu}=-g_{\mu\nu}\mathcal{L}_F+\frac{2\partial \mathcal{L}_F}{\partial g^{\mu\nu}}=F_{\mu\alpha}F_\nu^{\;\alpha}-\frac{1}{4}g_{\mu\nu}F_{\alpha\beta}F^ {\alpha\beta}
    \label{eq:55}
\end{equation}
\begin{equation}
    T_{\mu\nu}^\Phi=(\mathcal{D}_\mu\Phi)^*(\mathcal{D}_\nu\Phi)+(\mathcal{D}_\mu\Phi)(\mathcal{D}_\nu\Phi)^*-g_{\mu\nu}[(\mathcal{D}_\alpha\Phi)^*(\mathcal{D}_\beta\Phi)g^{\alpha\beta}-g_{\mu\nu}\lambda |\Phi|]
    \label{eq:56}
\end{equation}
Throughout the paper we assume vanishing electromagnetic strength tensor $F_{\mu\nu}=0$. For that purpose we could use the anzatz below:
\begin{equation}
    \Phi(t,r)=\phi(r)e^{i\omega t},\quad A_{\mu}(x^\mu)dx^\mu = A(r)dt
\end{equation}
Therefore, field equations for gauged boson Finch-Skea fluid sphere are expressed in the following form \cite{PhysRevD.88.024053}:
\begin{equation}
\kappa\rho= \left[e^{-\nu}
\left(\omega+qA\right)^2\right] \phi^2+\frac{ e^{-\lambda-\nu}
(A')^2}{2}+ \phi'^2e^{-\lambda}
\end{equation}
\begin{equation}
\kappa p_r =
\left[-e^{-\nu} \left(\omega+qA\right)^2\right]
\phi^2+\frac{e^{-\lambda-\nu}
(A')^2}{2}- \phi'^2e^{-\lambda}
\end{equation}
\begin{equation}
\kappa p_t =\left[-e^{-\nu} \left(\omega+qA\right)^2\right]
\phi^2-\frac{e^{-\lambda-\nu} (A')^2}{2}+ \phi'^2e^{-\lambda}
\end{equation}
Also, we could properly derive field equations for gauge (Maxwell) and scalar (Klein-Gordon) fields:
\begin{equation}
A''+\left( \frac{2}{r}-\frac{\nu'+\lambda'}{2}\right)A'-2 q
e^{\lambda} \phi^2 \left(\omega+qA\right)=0
\end{equation}
\begin{equation}
\phi''+\left( \frac{2}{r}+\frac{\nu'-\lambda'}{2}\right)\phi'+
e^{\lambda} \left[
\left(\omega+qA\right)^{2}e^{-\nu}\right]\phi=0
\end{equation}
As an initial conditions we have chosen the case with $\phi(0)=A(0)=const$ and $\phi'(0)=A'(0)=0$. In the next subsection we are going to match interior stellar spacetime and exterior Reissner-N\"ordrstrom one.

\subsection{Junction Conditions}
Since, we added minimally coupled $U(1)$ gauge field to our total Lagrangian, we need to redefine the junction conditions in the presence of interior charge. This time we will match interior and exterior spacetimes using the Reissner-N\" ordstrom metric tensor (which describes charged spherically symmetric vacuum), for which line element reads:
\begin{equation}
    ds^2 = \bigg(1-\frac{2M}{R}+\frac{Q^2}{R^2}\bigg)dt^2 - \bigg(1-\frac{2M}{R}+\frac{Q^2}{R^2}\bigg)^{-1}dr^2 - r^2 d\theta^2 - r^2 \sin^2 \theta d\phi^2
\end{equation}
Here, $Q$ is the total charge of the system, which is defined as
\begin{equation}
    Q=\int^R_0j^t\sqrt{-g}dr
\end{equation}
where $j^t$ is the only one non-vanishing component of four-current (Noether current) in the static spacetime. This current could be written as follows
\begin{equation}
    j^\mu = -ie[\Phi(\mathcal{D}^\mu\Phi)^*-\Phi^*(\mathcal{D}^\mu\Phi)]
\end{equation}
With the use of spherically symmetric line element of stellar spacetime we could rewrite the expression for total charge:
\begin{equation}
    Q=8 \pi q\int^{R}_{0}dr r^2
\left(\omega+qA\right)\phi^2e^{\frac{\lambda-\nu}{2}}
\end{equation}
To numerically solve field equations and coupled differential equations for scalar and gauge fields we need to make and assumption of value for charge $Q$. Since we are working in the RS spacetime, $Q<M$.
Using RS line element we could impose continuity conditions at the boundary:
\begin{equation}
    \mathrm{Continuity\;of\;}g_{tt}:\quad 1-\frac{2M}{R}+\frac{Q^2}{R^2}=e^{\nu(r)}
\end{equation}
\begin{equation}
    \mathrm{Continuity\;of\;}\frac{\partial g_{tt}}{\partial r}:\quad \frac{2M}{R^2}-\frac{2Q^2}{R^3}=-B \left(2 A \sqrt{C R^2}+B C R^3\right)
\end{equation}
\begin{equation}
    \mathrm{Continuity\;of\;}g_{rr}:\quad \bigg(1-\frac{2M}{R}+\frac{Q^2}{R^2}\bigg)^{-1}=e^{\lambda(r)}
\end{equation}
\begin{equation}
    \mathrm{Vacuum\;condition}:\quad p\rvert_{r=R}=0
\end{equation}
Solutions for above conditions are introduced below:
\begin{equation}
    \begin{gathered}
    A=\frac{\sqrt{R (R-2 M)+Q^2}}{R}-\frac{1}{2} B R \sqrt{C R^2}
    \end{gathered}
\end{equation}
\begin{equation}
    \begin{gathered}
    B=\frac{C \left(M R-Q^2\right)}{\left(C R^2\right)^{3/2} \sqrt{R (R-2 M)+Q^2}}
    \end{gathered}
\end{equation}
\begin{equation}
    \begin{gathered}
    C=\frac{1}{-2 M R+Q^2+R^2}-\frac{1}{R^2}
    \end{gathered}
\end{equation}
We will use these junction conditions to numerically investigate our model in the next subsections.

\subsection{Energy Conditions}
\begin{figure}[!htbp]
    \centering
    \includegraphics[width=\textwidth]{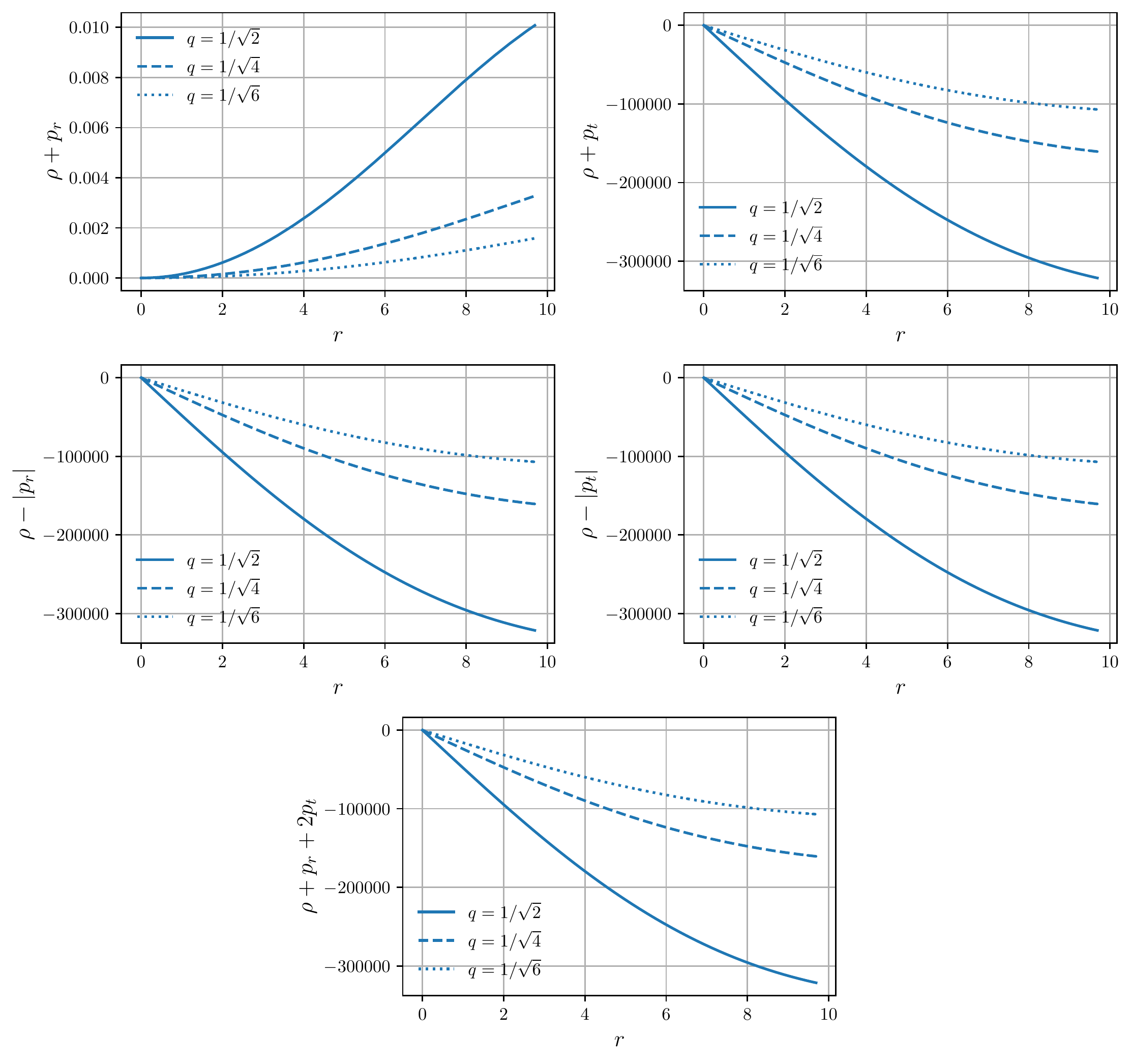}
    \caption{Null, Dominant and Strong energy conditions for Finch-Skea PSRJ1416-2230 star minimally coupled to $U(1)$ gauge field and massless complex scalar field. To numerically derive $T_{\mu\nu}$ components we use assumptions $Q=0.5$ and $\omega=10^{-6}$}
    \label{fig:13}
\end{figure}

Routinely, Null, Dominant and Strong energy conditions are plotted on the Figure (\ref{fig:13}). It is obvious that in our case only one energy condition, namely null energy condition for radial pressure is satisfied. Some of other violated EC's could be satisfied for bigger $\omega$ and different initial conditions for $A(r)$ and $\phi(r)$, but in that case $v^2>c^2$, which is unacceptable. Also, it is worth to notice that for Noether charge obeying inequality $|Q|<M$ if we will vary it's values, results will not change seriously and EC's will be still validated/violated.

\subsection{Gradients and Equation of State}
\begin{figure}[!htbp]
    \centering
    \includegraphics[width=\textwidth]{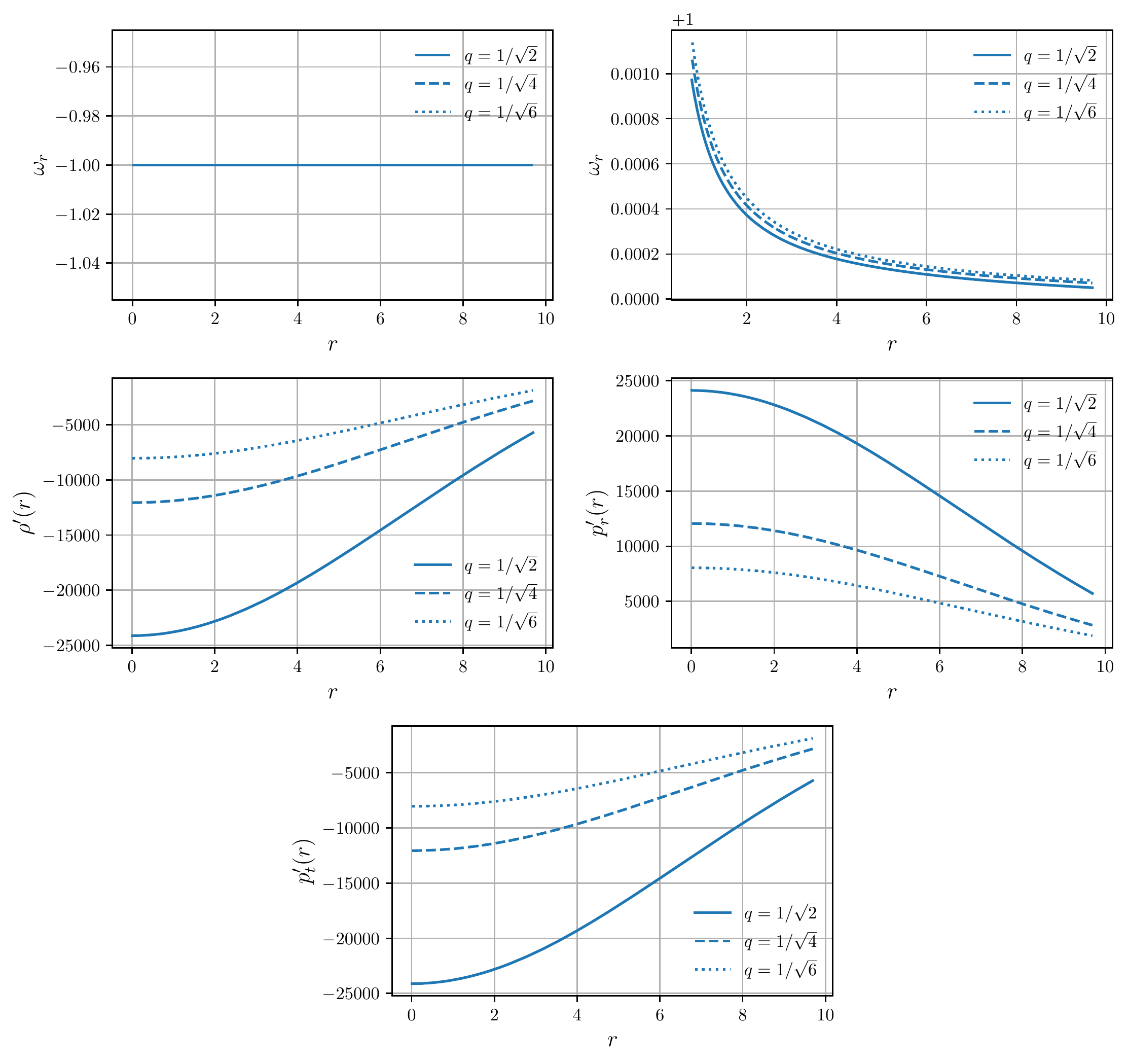}
    \caption{Equation of state and radial derivative of stress-energy tensor components for Finch-Skea star coupled to $U(1)$ gauge field and massless complex scalar field. To numerically derive that solutions we use assumptions $Q=0.5$ and $\omega=10^{-6}$}
    \label{fig:14}
\end{figure}

As usual, in this subsection we are going to study the radial derivatives of energy density, pressures and equation of state for Finch-Skea star minimally coupled to $U(1)$ gauge field, massless complex scalar field. Results of such investigation are illustrated on the Figure (\ref{fig:14}). From the plots we could conclude that radial fluid behave like dark energy and tangential like stiff fluid. On the other hand, gradients are positive for radial pressure and negative for energy density, tangential pressure.

\subsection{TOV stability}
\begin{figure}[!htbp]
    \centering
    \includegraphics[width=\textwidth]{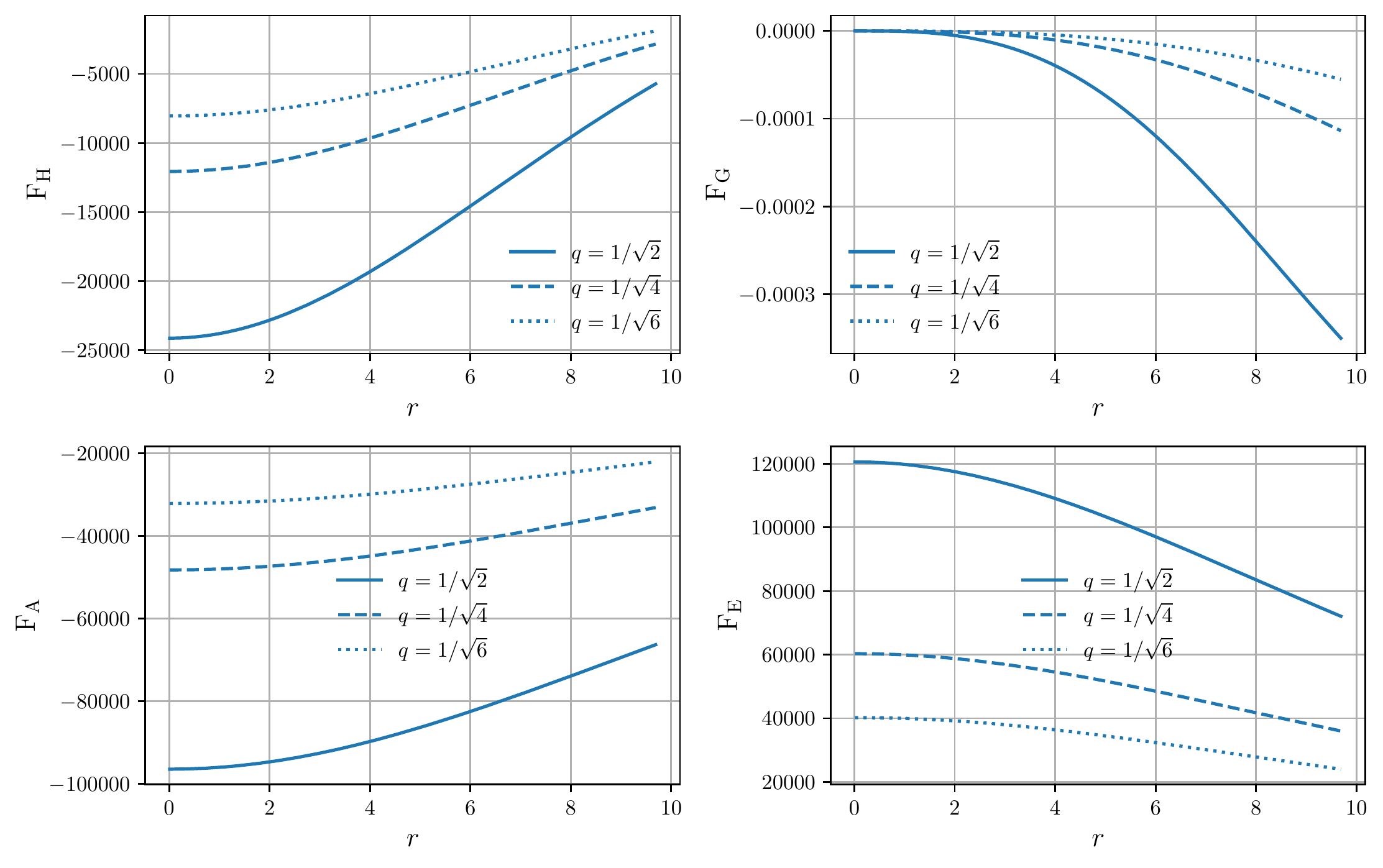}
    \caption{Forces present in the modified TOV equation for Finch-Skea star coupled to $U(1)$ gauge field and massless complex scalar field. To numerically derive that solutions we use assumptions $Q=0.5$ and $\omega=10^{-6}$}
    \label{fig:15}
\end{figure}

As well, we want to probe the stability of our stellar matter content through the well known modified TOV equation. We properly locate graphical results of TOV forces on the Figure (\ref{fig:15}). For particular case in the presence of gauge field, all of the forces except extra one are negative.

\subsection{Adiabatic index}
In this subsection we are going to study the stability of our stellar matter content from continuous adiabatic perturbations. Numerical solution for adiabatic index is located on the first plot of the Figure (\ref{fig:15}). As we see here, adiabatic index satisfy the $\Gamma>4/3$ constraint, so our stellar solutions in the presence of gauge field could be considered as stable from adiabatic perturbations.
\begin{figure}[!htbp]
    \centering
    \includegraphics[width=\textwidth]{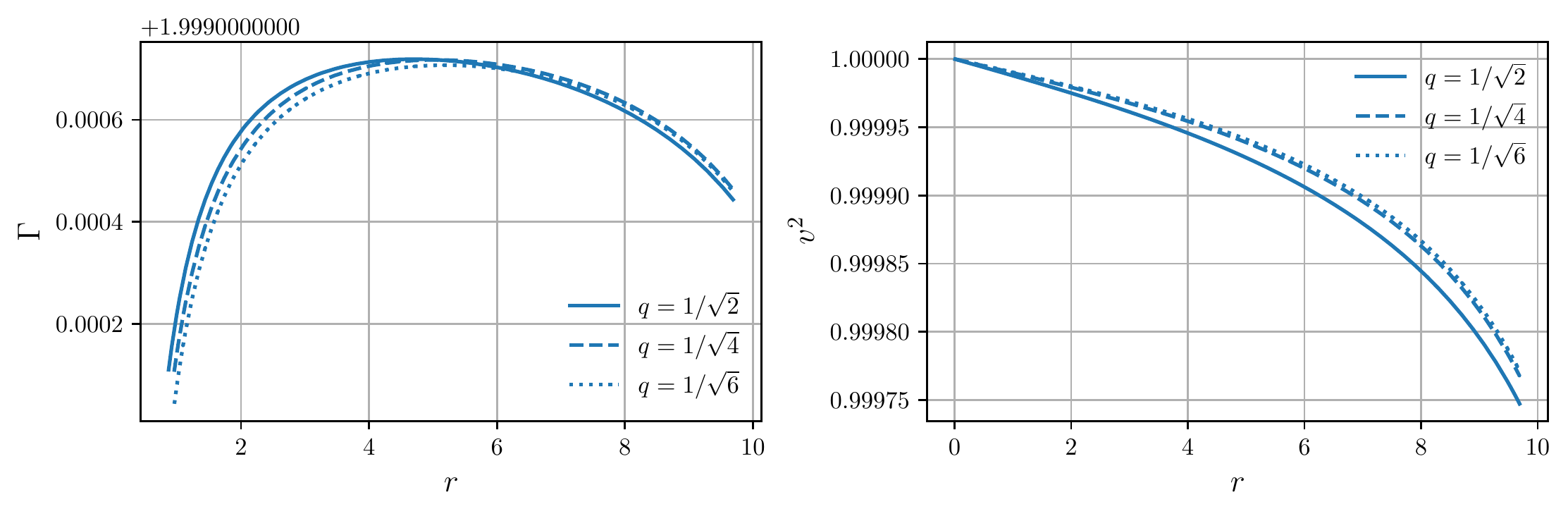}
    \caption{Anisotropic adiabatic index and speed of sound for Finch-Skea star coupled to $U(1)$ gauge field and massless complex scalar field. To numerically derive that solutions we use assumptions $Q=0.5$ and $\omega=10^{-6}$}
    \label{fig:16}
\end{figure}

\subsection{Speed of Sound}
Finally, we as well probed the speed of sound on the last plot of Figure (\ref{fig:16}). As we see, in our case fluid is ultrarelativistic, but does not exceed unity, which is required.

\subsection{Surface redshift}
\begin{figure}[!htbp]
    \centering
    \includegraphics[width=0.6\textwidth]{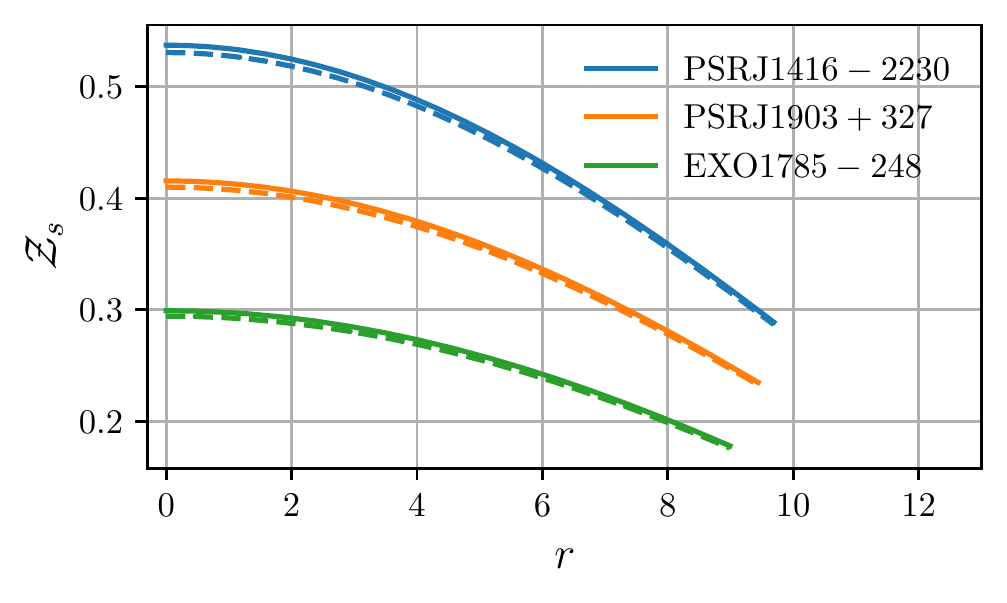}
    \caption{Surface (gravitational) redshift for gauged Finch-Skea stellar interior (parameter values are the same as for EC's). On the plot solid line represent the solution with $Q=0.9$ and dashed for $Q=1$}
    \label{fig:17}
\end{figure}

Since the junction conditions changed, surface redshift also will vary. Graphical representation of surface redshift is therefore placed on the Figure (\ref{fig:17}) for different values of total charge $Q$. As we noticed during the numerical analysis, surface redshift for charged stars is smaller that for uncharged ones, and consequently as $|Q|\to\infty$, $\mathcal{Z}_s\to 0$.

\section{Conclusions}\label{sec:5}
In this article, we investigated the spherically symmetric and static stellar solutions imposing Finch-Skea symmetry in the presence of such exotic matter fields as Bose-Einstein Condensate, Kalb-Ramond antisymmetric tensor field and gauge field with local $U(1)$ gauge symmetry. To numerically solve numerous equations and graphically plot the results of such investigation we have used compact star PSRJ1416-2230, which has the mass $M=1.69M_\odot$ and radius $R=9.69R_\odot$. In this section we are going to mention main results of our study:
\begin{itemize}
    \item \textbf{Energy Conditions}: for Bose-Einstein Condensate case Null and Strong energy conditions for both radial and tangential pressures were satisfied, so there is no exotic fluid in the stellar interior present. On the other hand, for the sake of causality every of the energy conditions mentioned in the paper were violated for Kalb-Ramond field and only radial NEC was validated for gauge field. We have already plotted energy conditions for each exotic matter field of our consideration on the Figures (\ref{fig:3}), (\ref{fig:8}) and (\ref{fig:13})
    \item \textbf{Equation of State}: for BE Condensate radial EoS parameter is regular and asymptotically $\phi'(0)\to-\infty$ describes Zeldovich fluid, tangential asymptotically describes $\Lambda$CDM dark energy like fluid. For Kalb-Ramond fluid situation differs, radial EoS is regular and vanish at the envelope, tangential one is negative at the stellar core and regular at the intermediate, envelope regions. Finally, for gauge field radial EoS describes DE-like fluid for any $|q|\in\mathbb{R}$ and tangential describes fluid with EoS which slightly deviates from the Zeldovich one and as $q\to0$, $\omega\to1$. For more detailed information and graphical representation of aforementioned results, refer to the Figures (\ref{fig:4}), (\ref{fig:9}) and (\ref{fig:14}) first row.
    \item \textbf{Gradients of perfect fluid stress-energy tensor components}: radial gradients were negative for energy density and radial pressure and positive for tangential one in the presence of BEC. For KB field everything is opposite, namely $\rho'(r)\land p_r'(r)\geq0$ and $p_t(r)\leq0$. Finally, for $U(1)$ gauge electromagnetic field energy density and tangential gradients were negative, radial was positive. Numerical evidence for provided statements could easily be found at the Figures (\ref{fig:4}), (\ref{fig:9}) and (\ref{fig:14})
    \item \textbf{TOV equilibrium}: we have as well probed the dynamical stability of our stellar object with minimally coupled exotic fields. For BEC as we noticed only hydrodynamical and extra forces were positive, the rest were negative and as well the vast contribution to the total TOV provided fluid anisotropy. On the other hand, for KB field there were only one negative force, namely hydrodynamical one, which was also the biggest one. Finally, for gauge field only extra force were positive. As usual, results are properly illustrated on the Figures (\ref{fig:5}), (\ref{fig:10}) and (\ref{fig:15})
    \item \textbf{Adiabatic index}: in the current study we also studied the stability of anisotropic matter inside compact star from adiabatic perturbations. As it was shown in this paper, $\Gamma>4/3$ for relatively big and negative initial condition for scalar field $\phi'(0)$ in the BEC case. Moreover, for KB case star is stable everywhere except the diverging regions. Ar final, for gauge field star is also stable everywhere. $\Gamma$ is plotted over the whole interior radial domain on the Figures (\ref{fig:76}), (\ref{fig:11}) and (\ref{fig:16})
    \item \textbf{Sound of speed}: the last quantity that we are going to discuss is the speed of sound. For object to obey causality condition, $v^2\leq c^2 =1$. For the BEC, KB and gauge fields this necessary condition was validated everywhere, as expected. Squared speed of sound is therefore plotted on the second plot of Figures (\ref{fig:76}), (\ref{fig:11}) and (\ref{fig:16})
\end{itemize}

Now, it will be handful to compare our results with the existing ones obtained for other compact star geometries. There were written numerous papers on compact stars within the General Theory of Relativity, however, here we are going to concentrate on the solutions, that are close to our own. For example, one could introduce Einstein-Klein-Gordon stars (compact stellar solutions endangered by scalar field), Einstein-Maxwell stars (analogically, stars endangered by Maxwell field that usually respects some gauge symmetries) or Boson/Proca stars as the closest "siblings" to our solutions.

For Einstein-Maxwell Buchdahl stars, good work in the GR theory is \cite{2021Ap&SS.366...26P}. In comparison to our results for stars with $U(1)$ background field, their solution respected all energy conditions. Moreover, because of the isotropy of their stellar solution, anisotropic forces in TOV vanished, hydrodynamical one were positive. Speed of sound squared had similar behaviour to ours, but the rate of change with $r$ was significantly bigger, EoS parameter was positive within the bounds of $\omega\in(0,1)$. Finally, the adiabatic index also had a lot bigger values, but inequality $\Gamma>4/3$ holds for both ours and their solutions.

On the other hand, bosonic stellar solutions unfortunately does not have such comprehensive analysis in the existing literature within General Theory of Relativity. However, there are some papers written within the Einstein-Scalar-Gauss-Bonnet (ESGB) theory, which is the viable theory of gravitation. For example, in the work of \cite{https://doi.org/10.48550/arxiv.2112.13391}, isotropic ESGB stellar solutions were probed. By comparing aforementioned model with our BEC stellar configuration, we could conclude that EC's for both models had practically the same behavior and asymptotically vanish. But in the EGSB case, $v^2$ and $\Gamma$ are monotonically decreasing functions while in our case they grow nearby stellar envelope. Moreover, for the case with Gauss-Bonnet corrections present, surface redshift were negative while in our case it was positive and decrease up to the envelope.

At the moment, there are no comprehensive studies (with the investigation of EC's, EoS, adiabatic and TOV stability) within the GR for such stellar configurations as Boson, Proca stars as well. However, from paper \cite{sym12122032} we could conclude that similar to our case, both Boson and Proca stars has positive energy density with the peak at the stellar origin and that $\rho\to0$ asymptotically with radial coordinate, which were observed for our case as well. Besides, there were carried out the investigation of magnetised BEC stars in \cite{Angulo:2022gpj}. In comparison to our BEC star solution, magnetised one had smaller surface redshift and both positive pressure and density with $\rho\gg p$, so that NEC/WEC/SEC and DEC are satisfied.

From the aforementioned results and graphical representations it is clear that our stellar solution is of non-singular nature and exhibits interesting properties in the physically interesting and viable models, such as GR expanded using exotic matter fields (BEC dark matter, KB field from string theory and gauge fields from electromagnetism). However, in the future studies it will be interesting to probe such model with Dirac spinor fields, QCD Axions and in the presence of non-homogeneous oscillating axion field with large scale correlations.

\section*{Data Availability Statement}
There are no new data associated with this article.

\section*{Acknowledgments}
PKS acknowledges National Board for Higher Mathematics (NBHM) under Department of Atomic Energy (DAE), Govt. of India for financial support to carry out the Research project No.: 02011/3/2022 NBHM(R.P.)/R\&D II/2152 Dt.14.02.2022. Sokoliuk O. performed the work in frame of the "Mathematical modeling in interdisciplinary research of processes and systems based on intelligent supercomputer, grid and cloud technologies" program of the NAS of Ukraine. We are very much grateful to the honorable referee and to the editor for the illuminating suggestions that have significantly improved our work in terms of research quality, and presentation.

\end{document}